\documentclass[twocolumn]{aastex631}

\usepackage{xcolor}
\usepackage{graphicx}
\usepackage{appendix}
\usepackage{amsmath,amssymb}
\usepackage{soul}
\usepackage{hyperref}

\received{Sep 30, 2024}
\revised{Jun 10, 2025}
\accepted{Jul 8, 2025}

\begin{document}

\title{ALMA reveals an Eccentricity Gradient in the Fomalhaut Debris Disk}

\correspondingauthor{Joshua Bennett Lovell}; \email{joshualovellastro@gmail.com}

\author[0000-0002-4248-5443]{Joshua Bennett Lovell}
\affiliation{Center for Astrophysics, Harvard \& Smithsonian, 60 Garden Street, Cambridge, MA 02138-1516, USA}

\author{Elliot M. Lynch}
\affiliation{Univ Lyon, Ens de Lyon, CNRS, Centre de Recherche Astrophysique de Lyon UMR5574, F-69230 Saint-Genis-Laval, France}

\author[0000-0002-4985-028X]{Jay Chittidi}
\affiliation{Department of Physics and Astronomy, Johns Hopkins University, Baltimore, MD 21218, USA}

\author[0000-0003-4623-1165]{Antranik A. Sefilian}
\affiliation{Department of Astronomy and Steward Observatory, University of Arizona, Tucson, AZ 85721, USA}

\author[0000-0003-2253-2270]{Sean M. Andrews}
\affiliation{Center for Astrophysics, Harvard \& Smithsonian, 60 Garden Street, Cambridge, MA 02138-1516, USA}

\author[0000-0001-6831-7547]{Grant M. Kennedy}
\affiliation{Department of Physics, University of Warwick, Coventry CV4 7AL, UK}

\author[0000-0001-7891-8143]{Meredith MacGregor}
\affiliation{Department of Physics and Astronomy, Johns Hopkins University, Baltimore, MD 21218, USA}

\author[0000-0003-1526-7587]{David J. Wilner}
\affiliation{Center for Astrophysics, Harvard \& Smithsonian, 60 Garden Street, Cambridge, MA 02138-1516, USA}

\author[0000-0001-9064-5598]{Mark C. Wyatt}
\affiliation{Institute of Astronomy, University of Cambridge, Madingley Road, Cambridge, CB3 0HA, UK}

\begin{abstract}
We present evidence of a negative eccentricity gradient in the debris disk of the nearby A-type main sequence star, Fomalhaut.
Fitting to the high-resolution, archival \textit{ALMA} 1.32\,mm continuum data for Fomalhaut (with a synthesised angular resolution of $0.76{\times}0.55''$; 4--6\,au), we present a model that describes the bulk properties of the disk (semi-major axis, width, and geometry) and its asymmetric morphology.
The best-fit model incorporates a forced eccentricity gradient that varies with semi-major axis, $e_f\propto a^{n_\mathrm{pow}}$, a generalized form of the parametric models of \citet{LynchLovell21}, with $n_\mathrm{pow}{=}{-1.75}{\pm}0.16$.
We show that this model is statistically preferred to models with constant forced and free eccentricities. 
In comparison to disk models with constant forced eccentricities, negative eccentricity gradient models broaden disk widths at pericenter versus apocenter, and increase disk surface densities at apocenter versus pericenter, both of which are seen in the Fomalhaut disk, and which we collectively term \textit{Eccentric Velocity Divergence}.
We propose single-planet architectures consistent with the model and investigate the stability of the disk over 440\,Myr to planet-disk interactions via N--body modeling.
We find that Fomalhaut's ring eccentricity plausibly formed during the protoplanetary disk stage, with subsequent planet-disk interactions responsible for carving the disk morphology. 
\end{abstract}



\section{Introduction}
\label{sec:intro}
Debris disks are cold (tens of K) dust belts formed via mutual planetesimal collisions \citep{Wyatt08, Matthews14, Hughes18, Marino2022}.
Whilst such collisions grind down disks over time, debris disks can survive well into (and beyond) the stellar main sequence, producing observable dust signatures over 10s--1000s of Myr.

Planet--disk interactions significantly alter the structures of debris disks, e.g., via depleting, over-populating, and warping planetesimal distributions.
On secular timescales, eccentric planets force their own eccentricities into the orbits of planetesimals via three-body interactions \citep[i.e., in a star--planet--planetesimal three-body system][]{Murray99, Mustill12}.
In resolved observations of disks, diagnostics such as the stellar offset from the disk center, relative brightness asymmetries, and relative width asymmetries have all been interpreted as metrics to derive disk eccentricities and eccentricity distributions \citep{Wyatt99, Pan16, LynchLovell21, LovellLynch2023}.

The most famous, and best-studied, eccentric debris disk, is that around the 7.66\,pc-distant, 440\,Myr old, A-type main sequence star, Fomalhaut \citep[derived from the Arabic name ``fam al-\d h\=ut (al-jan\=ub\=i)''  and translates to the ``mouth of the 
(southern) whale'';][]{Allen1963}.
Fomalhaut (HD~216956) is the brightest member of southern Piscis Austrinus constellation, though its debris has only been known of since 1985 \citep[via analysis of its Infrared Astronomical Satellite (IRAS) observations;][]{Aumann85}.
Since then, it has been resolved at higher-resolution with the Hubble Space Telescope \citep[HST; in optical scattered-light;][]{Kalas05, Kalas08, GasparRieke2020}, Spitzer Space Telescope and James Webb Space Telescope \citep[JWST; in near- and mid-infrared thermal emission;][]{Stapelfeldt+2004, Gaspar+2023}, with the Herschel Space Telescope \citep[in far-infrared thermal emission;][]{Acke+2012}, and the James Clerk Maxwell Telescope (JCMT) and Atacama Large sub-Millimeter Array \citep[\textit{ALMA}; in (sub-)millimeter thermal emission;][]{Holland+2003, Boley+2012, Matra15, MacGregor17, Matra17b}.
Most recently, both \citet{Kennedy20} and \citet{Chittidi_rev} presented evidence of a width asymmetry in the disk, with a broader pericenter width versus apocenter.
These observations definitively proved the eccentric nature of the debris disk: the star is offset from the geometric disk center (in both scattered light and thermal emission), and the disk presents wavelength--dependent disk ansae brightness asymmetries, specifically exhibiting `pericenter glow' in infrared emission \citep[see e.g.,][]{Wyatt99}, and `apocenter glow' in millimeter emission \citep[see e.g.,][]{Pan16}.
Both of these latter features are expected based on the structure and thermal conditions in Fomalhaut's disk at the observed resolution scales \citep{LynchLovell21}.

Here we present a millimeter model of Fomalhaut's main (outer) debris disk belt which incorporates a forced eccentricity gradient parameter, which we fit to the high-resolution \textit{ALMA} images of \citet{Chittidi_rev}. 
This model provides a novel interpretation of the disk's morphology; incorporating a steep, negative eccentricity gradient with semi-major axis.
In \S\ref{sec:obsmodel} we outline a physical description of the `eccentric velocity divergence' effect that describes this type of disk structure.
In \S\ref{sec:obs} we describe our model setup, fitting methodology, and results.
In \S\ref{sec:planets} we derive planet properties consistent with observations.
In \S\ref{sec:rebound} we consider the stability of these planetary scenarios with N-body modeling. 
We summarise our conclusions in \S\ref{sec:conclusions}.

\begin{figure*}
    \centering
    \includegraphics[width=1.0\textwidth]{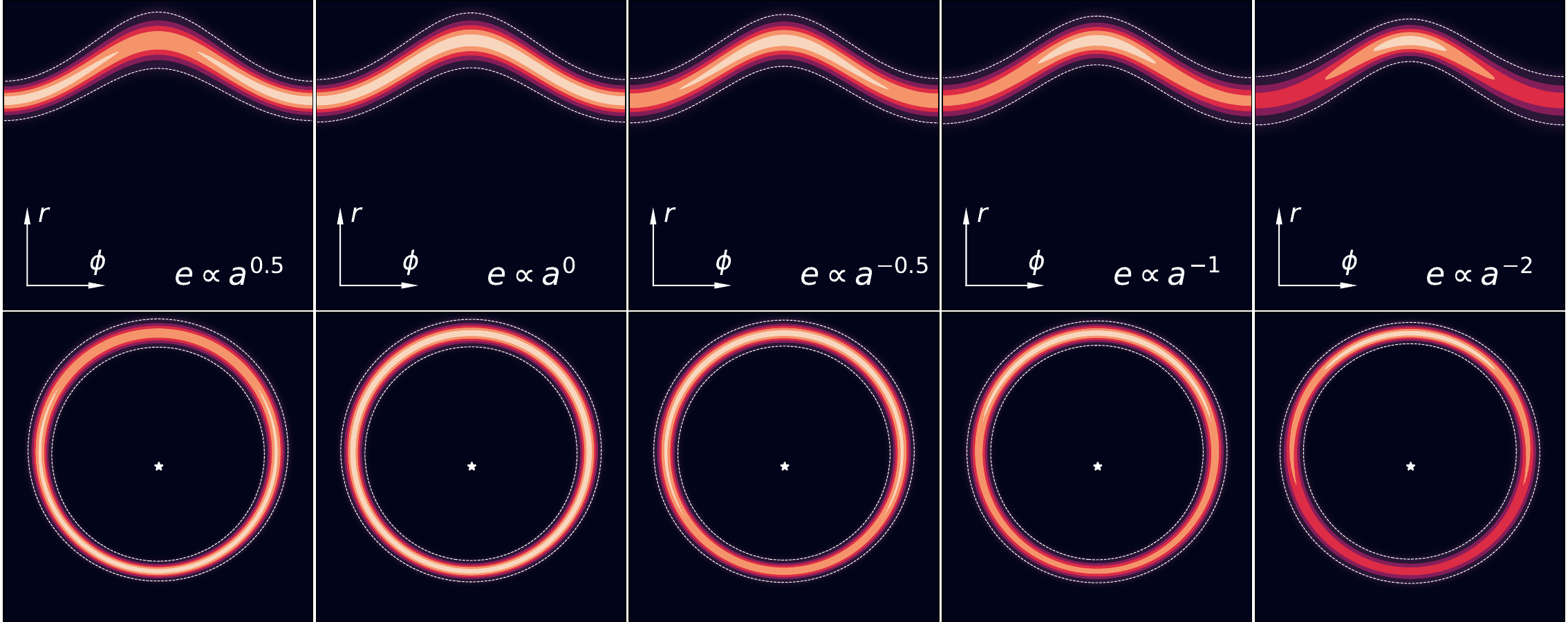}
    \caption{Normalized surface density maps (face-on projections, in (top) $r-\phi$ space, from $0-2\pi$ and $0-180$\,au, and (bottom) $x-y$ space) for different power-law eccentricity profiles (with their $e=e_f(a)$ functional form shown in the lower-right of each panel).
    The models all have apse-aligned argument's of pericenter, as well as radii, widths and eccentricities consistent with those of Fomalhaut, and an argument of pericenter directed south or towards 0 or 2$\pi$ in $r-\phi$ space.
    Contours in increasing brightness scale represent regions of higher surface density, set at the levels 10\%, 30\%, 50\%, 70\% and 90\% (10\% levels are shown with white-dotted contour lines).
    }
    \label{fig:gradient_example}
\end{figure*}

\section{Eccentric Velocity Divergence} \label{sec:obsmodel}
Central to the investigation that we outline in this work is the dependency between disk surface density distributions and their underlying eccentricity distributions.
This has been demonstrated recently in the literature by \citet{LynchLovell21} where simple power-law forced eccentricity profiles $e_f(a)\propto a^{n_{\rm{pow}}}$ with fixed ${n_{\rm{pow}}}$ values of $0$, $-1$, or $+1$ were modelled (where disk eccentricity gradients were either flat ($\partial e_f / \partial a = 0$), positive ($\partial e_f / \partial a >0$), or negative ($\partial e_f / \partial a <0$) respectively).
These models were shown to alter the resulting disk surface densities and emission morphologies. 
In this work we adopt the generalised form $e_f(a)\propto a^{n_{\rm{pow}}}$ (allowing for non-integer values) to examine power-law eccentricity profiles in the context of the Fomalhaut debris disk.
We illustrate the influence of this eccentricity distribution in Fig.~\ref{fig:gradient_example} showing normalized dust surface densities for a subset of power-law profiles ($0.5$, $0$, $-0.5$, $-1$, and $-2$, defined by $e_f(a)= e_{f,0}(a/a_0)^{n_{\rm{pow}}}$) adopting parameters consistent with the Fomalhaut disk (i.e., a semi-major axis of $a_0{=}139\,$au, a Gaussian disk width of $\sigma_r{=}15.6\,$au, a forced eccentricity $e_{f,0}{=}0.127$ at $a_0$), and an arbitrarily chosen $\omega_f{=}0.0$.
These plots show the same behaviour as described by \citet{LynchLovell21}, i.e., that surface densities in disks with positive eccentricity gradients are enhanced at pericenter, whereas disks with negative eccentricity gradients have surface densities raised at apocenter.
In addition, these models show that disk widths decrease where these are most dense, and broaden where these are least dense.
For example, negative eccentricity gradient profiles preferentially broaden disk widths at pericenter, and preferentially narrow disk widths at apocenter. 
To simplify discussion of the combined impact on disk surface densities and widths, we term this effect \textit{Eccentric Velocity Divergence}, which simply arises due to mass continuity in eccentric disks.

One can consider the effect on the disk width analytically by calculating the radial distance between two concentric orbits at pericenter ($\Delta r_{peri}$) and apocenter ($\Delta r_{apo}$).
In the case of fixed positive eccentricities, the definitions of the pericenter and apocenter radii result in widths $\Delta r_{peri} = \Delta a(1-e)$ and $\Delta r_{apo} = \Delta a(1+e)$, i.e., $\Delta r_{apo} > \Delta r_{peri}$ always for $e>0$.
However, for the general case of $e=e(a)$ (retaining only linear-order terms in the expansion of $e=e(a)$) the same derivation finds $\Delta r_{apo}\pm\Delta r_{peri} \approx \Delta a(1\pm (e + a\partial e /\partial a))$.
Consequently one can show that apocenter and pericenter widths are dependent on eccentricity gradients, with $\Delta r_{apo} > \Delta r_{peri}$ requiring $\partial e /\partial a\gtrsim -e/a$. 
For the case of $e_f(a)\propto a^{n_{\rm{pow}}}$, one finds that $\Delta r_{apo} > \Delta r_{peri}$ is only true if $n_{\rm{pow}}\gtrsim{-1}$, which can be seen by comparing the $e_f(a)\propto a^{-1}$ model (column four of Fig.~\ref{fig:gradient_example}) with other presented models.
Overall eccentric velocity divergence can induce similar features to those observed in the Fomalhaut debris disk, so we consider the possibility of its operation in this disk.

Power-law eccentricity profiles are physically motivated.
For example, in the idealised situation of an eccentric three-body system with either an \textit{inner} planetary perturber (i.e., a star--eccentric planet--planetesimal system), or an \textit{external} planetary perturber (i.e., a star--planetesimal--eccentric planet system) the disturbing function predicts analytical relationships between the semi-major axes and eccentricities of the perturbing planet and the planetesimal. 
To first--order (i.e., under the conditions of low eccentricity, and well-separated star--planet--(massless) planetesimals) these relationships can be derived as either $e_f(a) \propto a$ (for external planetary perturbations) or $e_f(a) \propto a^{-1}$ (for internal planetary perturbations).
These gradient slopes can steepen, for example, when the semi--major axis of a perturbing planet approaches those of its perturbers or if planetesimals are instead modelled with mass \citep[see e.g.,][]{Sefilian+2021, Sefilian2024}.
Indeed, such eccentricity distributions are present in the Kuiper Belt, whereby the distribution of trans-Neptunian Objects (TNOs) in the cold (classical) belt host steep $e_f(a)$ profiles \citep[e.g., at the level of a few percent per au;][]{Dawson12}.
Moreover, theoretical work on eccentric protoplanetary (gas) disks has shown that eccentricity gradients are necessary to avoid violent differential precession in $e_f(a)=\rm{const.}$ disks \citep{Teyssandier+2016}.
If debris disks form from primordial material with such eccentricity distributions, or are shaped by eccentric planet-disk interactions, it is plausible that their structures may be well described by the eccentric velocity divergence effect.

\section{Observations and modeling} \label{sec:obs}
\subsection{\textit{ALMA} observations} \label{sec:obsALMA}
For this study we fit to the images of Fomalhaut from \citet{Chittidi_rev}.
That work presents the calibration and data alignment procedure for the three \textit{ALMA} band~6 data sets included (P.I.: Aaron Boley ADS/JAO.ALMA\#2013.1.00486.S, P.I.: Paul Kalas ADS/JAO.ALMA\#2015.1.00966.S, and P.I.: Meredith MacGregor ADS/JAO.ALMA\#2017.1.01043.S) and utilizes standard {\tt CASA} ALMA reduction pipelines.
We use the same {\tt tclean} parameters for imaging as \citet{Chittidi_rev} with a briggs-weighting scheme and robust parameter of 0.5, which produces a mosaic image centred on the 2015 coordinates of Fomalhaut, with a pixel size of $0.075''$ and a synthesised clean beam with $0.76 \times 0.55''$, and beam position angle $-87.4^\circ$.
At the 7.66\,pc distance to Fomalhaut, this image provides a physical resolution of $6\times4\,$au.
We present the image in Fig.~\ref{fig:dataresiduals} (left).
The three observations that were combined to produce this image were all conducted with different mosaics, sensitivities, and \textit{ALMA} configurations (and thus effective angular resolutions). As a result, the final image has non-Gaussian noise properties, and primary beam attenuation that is dominated by the two-pointing mosaic of ADS/JAO.ALMA\#2017.1.01043.S, which had phase centers directed towards the disk ansae.
As can be seen in Fig.~\ref{fig:dataresiduals}, the data is therefore much noisier near to the minor axis of the disk in comparison to the two ansae.

\subsection{Model setup} \label{sec:obsmodelsetup}
We produce parametric models of optically thin disks consistent with those presented in \citet{Lovell21c}, \citet{LynchLovell21}, \citet{Lovell22} and \citet{LovellLynch2023}, i.e., by imaging 3-dimensional dust distributions within the {\tt RADMC-3D} framework \citep{Dullemond12}.
We fix the model wavelength to that of the band~6 observations ($\lambda=1.32\,$mm).
We fix the stellar parameters to values consistent with \citet{Mamajek2012}, i.e., the effective temperature ($T_\star=8500\,$K), stellar radius ($R_\star=1.8R_\odot$) and stellar mass ($M_\star=1.9M_\odot$), which together fully define the Kurucz stellar template spectra \citep[see][]{Kurucz79} from which the temperature conditions through the disk are derived by {\tt RADMC-3D}. 
The flux density at the location of the star (as adopted later) is a fixed value independent of these stellar parameters and the disk temperature conditions.
The models assume the dust to have a fixed size distribution between $0.9\,\mu$m--$1.0\,$cm with a $-3.5$ power-law index \citep[as per][]{Dohnanyi69}, and with grain densities of $\rho=3.5\,$g\,cm$^{-3}$.
We adopt a single Gaussian semi-major axis distribution, based on the (mean) disk semi-major axis ($a_0$) and Gaussian--width ($w_r$, which is related to the disk full width at half maximum (FWHM), by $\rm{FWHM}=2\sqrt{2\ln{2}}$\,$w_r$).
The surface density of the disk is modelled with a (mean) forced eccentricity ($e_{f,0}$, measured with respect to $a_0$), an argument of forced eccentricity ($\omega_f$, measured anti-clockwise from south), and a power-law index of $n_{\rm{pow}}$ (for a power-law eccentricity distribution of the form $e_f(a) = e_{f,0}( a/a_0)^{n_{\rm{pow}}}$), as per 
\begin{equation} 
\label{eq:1}
    \Sigma = \frac{1}{\sqrt{2\pi} w_r j} \exp \left( \frac{-(a-a_0)^2}{2w_r^2} \right)
\end{equation}
with $j$ the coordinate system Jacobian determinant \citep[as also presented as Equation~8 in][]{LynchLovell21}
\begin{equation} \label{eq:2}
    j = \frac{1-e(e+a \partial e /\partial a)}{\sqrt{1-e^2}} (1 - q \cos{E})
\end{equation}
for the eccentric anomaly ($E$)
\begin{equation} \label{eq:3}
\cos{E} = \frac{\cos{(\phi-\omega_f)} + e}{1 + e \cos{(\phi-\omega_f})}
\end{equation}
and the orbital intersection parameter ($q$, in the case of an untwisted disk, i.e., $\partial \omega_f/\partial a=0$)
\begin{equation} \label{eq:4}
q = \frac{a \partial e/\partial a}{1-e(e + a \partial e/\partial a)}.
\end{equation}
To avoid orbital intersections we require $|q|<1$.
Equation~\ref{eq:1} fully describes the plots presented in Fig.~\ref{fig:gradient_example} (gridded on ${r-\phi}$ and ${x-y}$ space)\footnote{The code for generating eccentric ring models and reproducing these plots is available via DOI: \url{https://doi.org/10.5281/zenodo.16758671} and at  \url{https://github.com/astroJLovell/eccentricDiskModels}.}.

We adopt a single Gaussian--distributed \textit{vertical} density profile, with a vertical aspect ratio ($h=H/r$) for $H$ the classical scale-height for a given disk radius $r$, such that the vertical linear density is
\begin{equation} \label{eq:5}
    l_\rho = \frac{1}{\sqrt{2\pi}H} \exp \left( \frac{-z^2}{2H^2} \right).
\end{equation}
This parametrisation defines the total density  $\rho=M_{\rm{dust}}\Sigma l_\rho$, where we introduce a dust mass scaling parameter ($M_{\rm{dust}}$) which uniformly scales model pixel intensities to mass units.

\begin{figure*}
    \centering
    \includegraphics[width=1.0\textwidth,trim={0cm 0.0cm 0cm 0cm},clip]{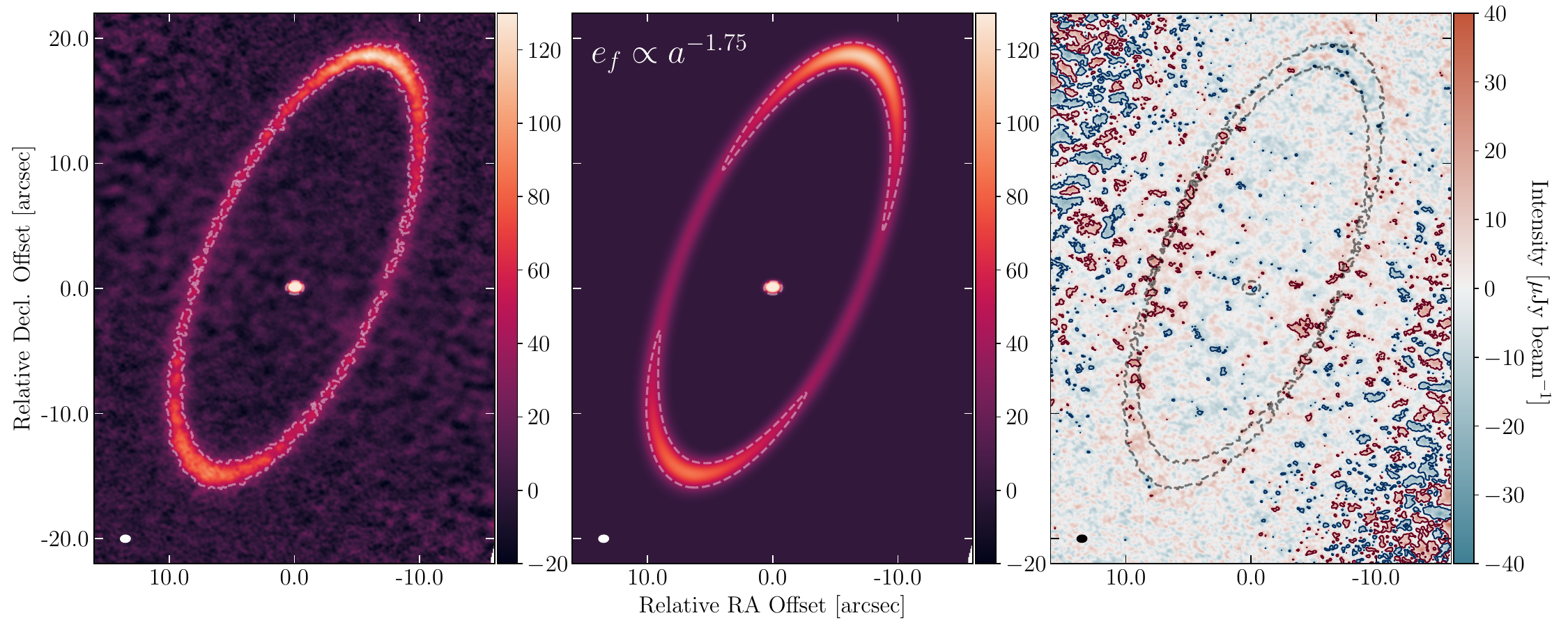}
    \caption{Left: \textit{ALMA} data as presented in \citet{Chittidi_rev} which we fit to in this work. Center: Best--fit $e_f(a)\propto a^{-1.75}$ model (on same image scale). Right: residual emission after we subtract our best-fit model from the \textit{ALMA} data (left). Contours are shown at the $\pm3\sigma$ level in the residuals, and at the $5\sigma$ level for the data/model. Beams are shown in the lower--left of each plot. Emission remains present in the minor axis region of the residuals, however this is dominated by imaging noise (enhanced due to the primary beam response) and background sources (such as `GDC', see \citet{Gaspar+2023} and \citet{KennedyLovell2023}). North is up, east is left.}
    \label{fig:dataresiduals}
\end{figure*}

By default our models are face-on, oriented north-south (with the pericenter direction southwards, apocenter direction northwards, as per Fig.~\ref{fig:gradient_example}).
We use {\tt RADMC-3D} to project and image models on the sky-plane.
We provide parameters for the position angle ($\rm{PA}$, which rotates our model anti-clockwise referenced to north), inclination ($i$, which inclines our model with reference to angles between $i=0^\circ$ for face-on projections, and $i=90^\circ$ for edge-on projections), and offsets in the right ascension ($\Delta \rm{RA}$) and declination ($\Delta \rm{Dec}$) respectively (applied uniformly to both the stellar location and disk focus). 
We present in Fig.~\ref{fig:dataresiduals} (middle) one such imaged model with Fomalhaut's disk and stellar parameters (convolved with the same clean beam parameters as the imaged \textit{ALMA} data, using the best-fit parameters inferred later).

\subsection{Model fitting} \label{sec:obsmodelfitting}
Our fitting approach is consistent with the procedures of \citet{Lovell21c, Lovell22}, except in this work we fit in the image domain instead of the visibilities\footnote{It is in principle possible to fit such models in the visibility domain, however this demands a much greater computational time investment. We note that model convergence can be achieved with this image-based setup in a modest 1--2 days, whereas in the visibility domain a much more intensive 1--2 weeks per model is required \citep[see][]{Chittidi_rev}, and thus the approach here enables us to investigate a broad array of model setups.}.
We fit our model to the \textit{ALMA} data with {\tt emcee}, an `Affine Invariant Markov Chain Monte Carlo (MCMC) Ensemble Sampler'\footnote{I.e., see:  \url{https://emcee.readthedocs.io/en/stable/}} \citep{Goodman10, FM13}.
Samples of the posterior distribution (generated by {\tt emcee}) are determined from the assumed prior distributions and a Gaussian likelihood function ($\propto \exp({- \chi^2/2})$), with
\begin{equation}
    \chi^2 = \sum_{i,j=1}^{\rm{N_{pix}}} \frac{|| I_{\rm{data,i,j}} - I_{\rm{model,i,j}} ||^2}{\delta I_{\rm{data}}^2}\frac{l_{\rm{pix}}^2}{\Omega_{\rm{beam}}}\mathcal{M}_{\rm{i,j}},
    \label{eq:chi2}
\end{equation}
for an image with size $\rm{N_{pix}}\times\rm{N_{pix}}$, intensity in each pixel ($I$, for either the model or data), the image noise ($\delta I_{\rm{data}}$, see further below), the beam size ($\Omega_{\rm{beam}}$), image pixel size ($l_{\rm{pix}}$, where the latter term $l_{\rm{pix}}^2/\Omega_{\rm{beam}}$ re-weights $\chi^2$ to account for pixel-pixel correlation), and the fitting mask ($\mathcal{M}_{\rm{i,j}}$).
We fit to the primary-beam corrected image ($I_{\rm{data,i,j}}$) and measure the $\chi^2$ in the residual map corrected by the primary beam response, assuming the {\tt CASA} form of ALMA's primary beam (as described in e.g. \href{https://help.almascience.org/kb/articles/pdf/how-do-i-model-the-alma-primary-beam-and-how-can-i-use-that-model-to-obtain-the-sensitivity-pr}{ALMA Knowledgebase, Mar 2025}).
The \textit{ALMA} image has a non-Gaussian noise distribution most prominent near to the disk minor axis given the image derives from different mosaic patterns from different ALMA configurations (see Fig.~\ref{fig:dataresiduals}).
By comparing fits to data with different masking conditions, we proceeded to fit with a mask defined by a cut-off in the \textit{ALMA} imaged primary beam (only fitting emission where the primary beam response exceeds $0.66$), and a projected radius cut-off (only fitting to emission within a projected $200$\,au distance from the stellar center).
We present in Appendix~\ref{appX} (Fig.~\ref{fig:mask}) the location of the fitting mask and the primary beam response for the data as well as the discussion of and results of fitting the General Model to the data without including the primary beam cut-off in the map.
The mask simply multiplies pixel values with either $\mathcal{M}_{\rm{i,j}}=1$ or $\mathcal{M}_{\rm{i,j}}=0$ and is applied to both the data and model identically.
The mask choice focuses the fitting on high SNR disk emission, avoiding imaging artifacts and background sources near the disk semi-minor axes \citep[e.g., such as `GDC' (the Great Dust Cloud) as discussed by][]{KennedyLovell2023}.
We note that by masking in the fitting stage rather than during model generation, our models still comprise azimuthally-complete disks.

We adopt an uncertainty of $\delta I_{\rm{data}}=8.0\,\mu$Jy\,beam$^{-1}$ based on the root-mean-square of pixel intensities in disk-free regions of the \textit{ALMA} image which is scaled by the primary beam response during fitting, down-weighting noisier regions of the data.
We find from test iterations of fitting models to the data with {\tt emcee} that the residual maps have standard deviation ${\approx}8.0\,\mu$Jy\,beam$^{-1}$ (i.e., as expected given the image data) and mean ${<}1.0\,\mu$Jy\,beam$^{-1}$ (i.e., very close to 0, as expected for well-fitted models).
To verify the fitting methodology, we tested this procedure on the cycle~3-only \textit{ALMA} data of Fomalhaut as described in Appendix~\ref{appB}.

\subsection{Modeling results} \label{sec:results}
\subsubsection{Fitting an EVD disk model} \label{sec:evdresults}
We fit the stellar flux and location separately to the disk to reduce the number of free parameters in our fitting (since we simply fit the stellar \textit{flux}, we leave fixed the physical stellar inputs that derive the disk temperature conditions, e.g., $R_\star$, $M_\star$ and $T_\star$).
We adopt a small aperture mask over the full image to fit the stellar flux and location, masking all but a $10\times10$ pixel box.
We find that the star is best-fitted with offsets consistent with 0.0 in both axes ($\Delta \rm{RA}=0.01{\pm}0.04''$ and $\Delta \rm{Dec}=0.01{\pm}0.04''$, in accordance with the realignment and imaging procedure of \citet{Chittidi_rev}, and a flux of $0.737\,$mJy (i.e., $F_{\rm{\star}}{=}0.74{\pm}0.08\,$mJy accounting for a 10\% \textit{ALMA} flux calibration uncertainty in quadrature with our fitting error).

We run an initial fitting routine with {\tt emcee} (with 48 walkers, 3000 steps) to estimate the convergence timescale and to further verify the fitting results for the higher resolution data.
This initial fitting routine showed several thousand more steps were required to achieve convergence, which we discuss in detail in Appendix~\ref{appC}.
Nevertheless, the physical model parameters we measured were all consistent with previous works. Of note, we fitted $\rm{PA}=336.19{\pm}0.10^\circ$, a (Gaussian) vertical aspect ratio of $h=0.0157{\pm}0.0013$ and $\omega_f=15.6{\pm}0.6^\circ$.
The Gaussian vertical aspect ratio is slightly lower than the 0.02 assumed in \citet{MacGregor17}, and consistent with that fitted by \citet{Kennedy20}\footnote{Although we note that the upper limits on inclination in \citet{Kennedy20} were not as stringent as they should have been given the data. 
We have traced this to a coding error that resulted in the inclination parameters only affecting radial structure (i.e., the disk models were flat). With the vertical structure parameterisation correctly included, we re-ran those models and instead yield an upper limit on the Gaussian vertical aspect ratio of $h<0.03$, with only very small differences for other parameters (e.g. $e_{f} = 0.126 \pm 0.001$ compared to $0.125 \pm 0.001$ previously).}.
Given the resolution of the \textit{ALMA} data, the low value that we determine for $h$, and the low $h$ constrained by previous work, we note the vertical structure is plausibly still unresolved.
Further, in \citet{Kennedy20} the disk rotation was modelled with the longitude of ascending node $\Omega=156.4^\circ$ (measured anti-clockwise from north) which is directed towards the south-west ansa, whereas we model $\rm{PA}=336.19^\circ$ anti-clockwise from north to align with the north-east ansa (i.e., these are simply $180^\circ$ offset, and thus fully consistent).
Further, the definition of $\omega'_f$ of \citet{Kennedy20} is measured with respect to the longitude of ascending node, whereas the definition we adopt is with respect to south, i.e., \textit{independent} of the $\rm{PA}$ (with both again measured anti-clockwise).
As such the value of $\omega'_f=41\pm1^\circ$ reported in \citet{Kennedy20} would have been measured as $\omega_f = \Omega -\pi + \omega'_f$, i.e., $17.1{\pm}1.0^\circ$ via the model setup we describe here.
In this work, we consistently fit values within a few degrees of this value with similar uncertainties, and thus we deem these to likewise agree. 

To reduce the number of parameters we fit to, we proceeded by fixing $\Delta \rm{RA}$ and $\Delta \rm{Dec}$, and fitted only for the disk mass ($M_{\rm{disk}}$), semi-major axis ($a_0$), disk width ($w_r$), (mean) forced eccentricity ($e_{f,0}$), argument of forced eccentricity ($\omega_f$), the power-law index for the forced eccentricity gradient ($n_{\rm{pow}}$), the inclination ($i$), the position angle ($\rm{PA}$), and the (Gaussian) vertical aspect ratio ($h$).
We term this the `General' model. 
We present in Appendix~\ref{appA} the posterior distribution of the converged {\tt emcee} chains, and discuss its convergence criteria in Appendix~\ref{appC} after re-running the fitting routine for 6000 steps.
We find a significant determination of the forced eccentricity gradient power-law index ($n_{\rm{pow}}=-1.75\pm0.16$), and parameters for the disk mass, semi-major axis, width, eccentricity, argument of pericenter, inclination, position angle and vertical aspect ratio that are all consistent with \citet{MacGregor17} and (where presented) \citet{Kennedy20}.

We assess the properties of our residual emission map to investigate if our model is a good fit.
For the complete masked region, we measure a pixel variance equal to the square of our measured $\sigma_\mathrm{image}$, a pixel mean equal to 0.1$\times$$\sigma_\mathrm{image}$, and a distribution that follows a Gaussian-distribution.
These values imply that only minimal \textit{excess} emission remains in the final residual image (which may be fully explained by the presence of bright point sources, that may be background sources), and that it varies as would be expected if it is populated only by noise.
The residual map as shown in Fig.~\ref{fig:dataresiduals}  appears to demonstrate the the General model is a very good fit to the data.
For example, in the ansae the broader and fainter pericenter are fitted within the noise, and in the apocenter, the peak is reproduced within the noise.
In the residual maps at the two ansae, we see only small (sub-beam sized) residuals that exceed $3\sigma$ (none of which ${\gtrsim}4\sigma$).
Along the minor axes, the model appears to be a worse fit, given the model is fainter than the data in those locations.
Some of these larger peaks are either background sources \citep[i.e., `GDC' as noted by][]{KennedyLovell2023}, or noisier regions of the data.
The residual emission coincident with the apocenter outer edge appears to be slightly over-subtracted by our model, which suggests that our width may not be quite narrow enough in this location\footnote{To investigate this, we attempted a more complex radial profile fit to the data, allowing distinct $w_r$ values internal/external to the semi-major axis in case these residuals resulted from fitting a Gaussian surface density profile to data that may not be Gaussian in emission. Whilst the fit yielded improved residuals (with inner-edge and outer-edge widths of $w_{\rm{r,\,in}}=11.2{\pm}0.7$au and $w_{\rm{r,\,in}}=15.7{\pm}0.7$au respectively, a smaller mean semi-major axis of $a=137.6{\pm}0.4$au and otherwise statistically consistent best-fit model parameters) this model neither fully removed these negative residual features, nor was the improvement in $\chi^2$ statistically significant versus the simpler Gaussian model.}.
These residual features may therefore suggest that a steeper eccentricity gradient model may be needed, but the existing data does not prefer such a model.

Similar \textit{negative} apocenter residuals are present in the residual maps of \citet{Kennedy20} and also the fit we performed to the lower resolution data (discussed in Appendix~\ref{appB}).
However, we note that the residuals of \citet{Kennedy20} host in addition a strong \textit{positive} residual (located in-between the negative residual regions), i.e., those models cannot account for the apocenter brightness enhancement in the disk.
In contrast we find no positive residuals in the disk ansae, in our fits to \textit{both} the high and low resolution data.
Our findings thus present evidence that the Fomalhaut debris disk can be understood in the paradigm of eccentric velocity divergence, with this hosting the first evidence of an eccentricity gradient in its debris belt. 
We collate our best-fit parameter results in Table~\ref{tab:results} for all our fitting setups. 
We investigate in \S\ref{sec:efree} whether the data prefers a model with free eccentricity and either a constant or negative forced eccentricity profile. 

\begin{table*}[]
    \centering
    \caption{Best--fit model results for the fits to the high--resolution {\it ALMA} data (top) and the best-fit `Verify' model fit to the lower-resolution data (bottom). The $\Delta\chi^2$ values are all with reference to the General model. Values presented in square brackets were fixed to the presented value during fitting.
    For completeness we also present the parameters determined for the verification model which only used the low-resolution data of \citet{Kennedy20}.}
    \begin{tabular}{cc|ccc|c}
    \hline
    \hline
    Parameter & Units & General & Proper 1 & Proper 2 & Verify \\
    \hline
    $M_{\rm{dust}}$ &$10^{-2}\,M_\oplus$ & $18.70{\pm}0.20$ & $20.90{\pm}1.3$ & $21.70{\pm}1.3$ & $19.90{\pm}0.20$\\
    $a_0$ &au & $138.79{\pm}0.12$ & 
$138.93{\pm}0.12$ & $138.84{\pm}0.12$ &$138.89{\pm}0.10$ \\
    $w_r$ &au & $13.51{\pm}0.29$ & $9.3{\pm}0.6$ &$12.8{\pm}1.0$& $15.3{\pm}0.3$ \\
    $e_{f,0}$& \% & $12.56{\pm}0.12$ & $12.56{\pm}0.11$ &$12.57{\pm}0.12$& $13.46{\pm}0.14$ \\ 
    $e_p$ &\% & [$0.0$] & $3.89{\pm}0.16$ & ${<}3.6$& [$0.0$] \\ 
    $\omega_f$ &deg & $15.2{\pm}0.6$ & $15.0{\pm}0.6$ &$15.2{\pm}0.6$& $19.1{\pm}0.6$ \\ 
    $n_{\rm{pow}}$ &--& $-1.75{\pm}0.16$ & [$0.0$] & [$-1.0$]& $-1.16{\pm}0.16$ \\
    $h$& \% & $1.57{\pm}0.13$ & $1.56{\pm}0.14$ &$1.50{\pm}0.14$& [1.43]\\
    $i$& deg & $66.44{\pm}0.09$ & $66.44{\pm}0.09$ & $66.43{\pm}0.09$& [66.6]\\
    PA &deg & $336.19{\pm}0.06$ & $336.22{\pm}0.05$ &$336.20{\pm}0.06$& [336.4] \\
    \hline
    $\Delta\chi^2$ &--&0.0&$+70.0$&$+21.5$& --\\ 
    \hline
    \hline
    \end{tabular}
    \label{tab:results}
\end{table*}

\subsubsection{Sensitivity to free eccentricity?} \label{sec:efree}
To test the possibility that the power--law eccentricity distribution model fit is biased by not accounting for the `free' (or `proper') eccentricity parameter ($e_p$), we produce a disk model which includes, $e_p$, $e_f$ and a power--law distribution in the forced eccentricity.
We solve analytically the orbit equation for the case of a disk with both free and forced eccentricities following $e\propto a^{n_{\rm{pow}}}$ with fixed values of ${n_{\rm{pow}}}=0$, $-1$ (and leave to future work the general case for any value of ${n_{\rm{pow}}}$, i.e., for $e = e_{f,0}(a/a_0)^{n_{\rm{pow}}}\exp{[i(\phi - \omega_f)]} + e_p\exp{[i(\phi - \omega_p])}$, for a system with (mean) forced and free eccentricities ($e_{f,0}$ and $e_p$ respectively) and arguments of forced and free eccentricities ($\omega_f$ and $\omega_p$ respectively).
To model the free eccentricity, we adopt the same approach as \citet{LovellLynch2023}, i.e., we linearly sum independent `families' of orbits (each comprising a single argument of free eccentricity $\omega_p$) uniformly over the range of $0{\leq}\omega_p{<}2\pi$, where each family comprises a disk with $M_{\rm{disk}}$, $a_0$, $w_r$, $e_{f,0}$, $\omega_f$, $e_p$, and (fixed) $n_{\rm{pow}}$.
\citet{LovellLynch2023} demonstrated that this parametric modeling approach for the free eccentricity distribution yields consistent results to methods that instead model individual particles (such as \citet{MacGregor17}, \citet{Kennedy20}, where $n_{\rm{pow}}=0$ was implicitly adopted).
Given the negative best--fit $n_{\rm{pow}}$ that we earlier inferred (and that previous modeling efforts implicitly had $n_{\rm{pow}}=0$) we consider here only the two cases of $n_{\rm{pow}}{=}0$ and $n_{\rm{pow}}{=}{-1}$.

\begin{figure*}
    \centering
   \includegraphics[width=1.0\textwidth, trim={0cm 1.0cm 0cm 0cm},clip]{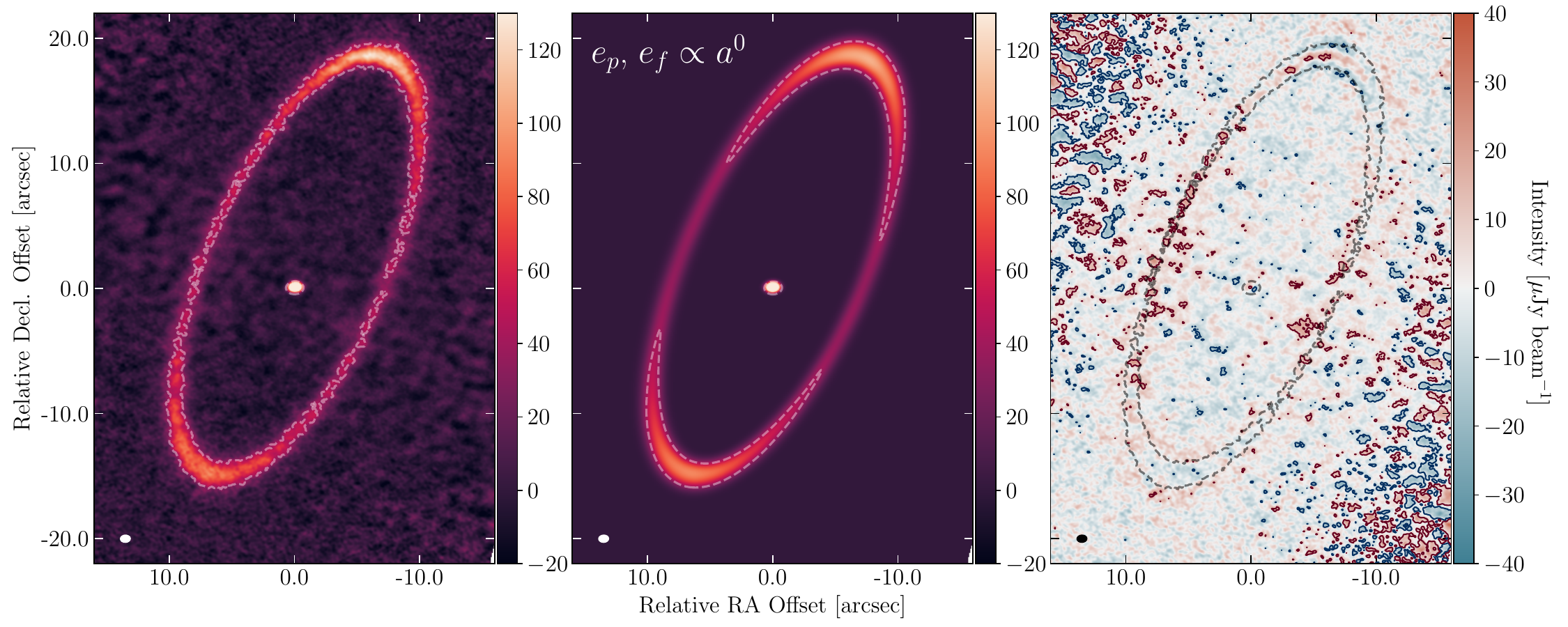}
    \includegraphics[width=1.0\textwidth]{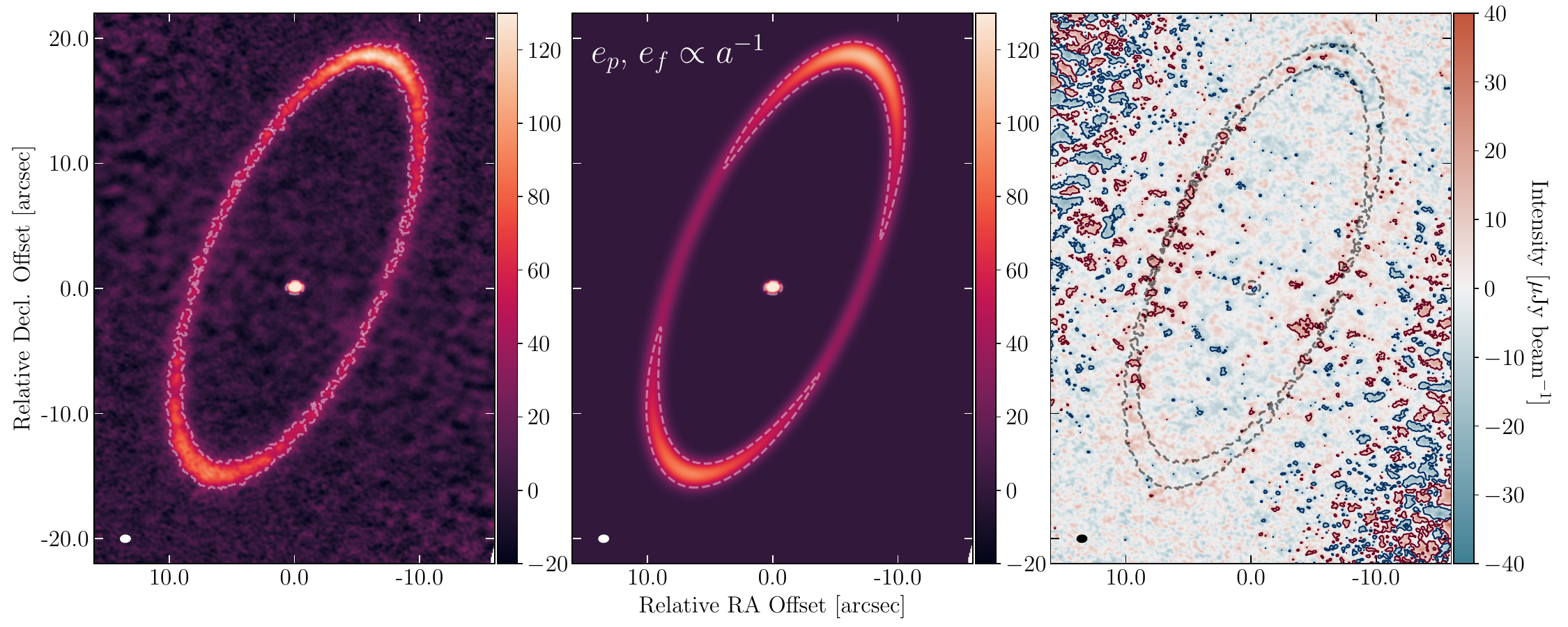}
        \caption{Left: \textit{ALMA} data. Center: Best--fit models  which include $e_p$ (on same image scale, with top for the model fixed with $n_{\rm{pow}}{=}0$, and $n_{\rm{pow}}{=}{-1}$ for bottom). Right: residual emission after we subtract the best-fit models from the \textit{ALMA} data respectively. 
    Contours are shown at the $\pm3\sigma$ level in the residuals, and at the $5\sigma$ level for the data/model. Beams are shown in the lower--left of each plot. In all plots, north is up, east is left.}
    \label{fig:residseFree}
\end{figure*}

Firstly, by fitting a 6--parameter model (i.e., with $n_{\rm{pow}}$ fixed to $0$) and $e_p$ allowed to freely vary, we run the same model fitting procedure described earlier with {\tt emcee} (for 3000 steps) and find that the best--fit solution results in $e_p{=}3.85{\pm}0.17\%$, i.e., this requires a significant free eccentricity. 
We term this model `Proper 1'.
Secondly, re-running this process instead with $n_{\rm{pow}}$ fixed to $-1$ (for 5000 steps to ensure convergence), we find that the best--fit solution results in $e_p{=}1.9{\pm}0.9\%$ (for which we place an upper--limit on $e_p{<}3.59\%$ based on the 99.7th-percentile of all chains after burn-in).  
We term this model `Proper 2'.
We present the model and residuals of these fitting processes in Fig.~\ref{fig:residseFree}, all best--fit values in Table~\ref{tab:results}, and the corner plots for the posteriors in Figs.~\ref{fig:corners_ep1} and \ref{fig:corners_ep2}.
One can see that whilst the $n_{\rm{pow}}=-1$ provides a reasonable fit (albeit poorer than the model with a steeper eccentricity gradient and no free eccentricity) the model with $n_{\rm{pow}}=0$ and a significant free eccentricity is poorer at interpreting the broader width at pericenter and the brightness enhancement at apocenter consistent with the findings in \citet{Chittidi_rev}, and also those of \citet{MacGregor17} and \citet{Kennedy20}.
We note that by definition there is a degeneracy between $e_p$ and $w_r$ which results in the wide range of $w_r$ posterior distributions.

\subsubsection{Which model is preferred?}
In Table~\ref{tab:results} we define $\Delta\chi^2$ as the relative difference in the best-fit value $\chi^2$ of the `General' model versus the models with free eccentricities, where the General model returns a lower $\chi^2$ than the Proper 1 and 2 models by 70 and 22 respectively.
We quantify these differences statistically by estimating the associated probability that (given each of the models has 9 free parameters) the relative difference in $\chi^2$ is significantly improved by calculating the Incomplete Gamma function for each value of $\Delta\chi^2$, i.e.
\begin{equation}
    P = \frac{1}{\Gamma(N_{\rm{par}}/2)} \int^{\Delta\chi^2}_0 t^{\frac{N_{\rm{par}}}{2} - 1}\exp{-t}dt \,\,.
\end{equation}
In terms of confidence intervals, these respective $\Delta\chi^2$ values therefore imply that the General model is favored over the Proper 1 and 2 models by $6.7\sigma$ and $2.6\sigma$ respectively, and that the Proper 2 model is favored over Proper 1 by $5.2\sigma$.
We note that in a third version of the Proper model which instead had a fixed eccentricity power-law gradient of $-2$, we found the General model still had a lower $\chi^2$ by 10.6, and thus performed statistically as well, with a preference of only $1.0\sigma$. 
Overall, these differences imply that whilst the disk may host some small amount of free eccentricity, the eccentric disk is dominated by the presence of a forced eccentricity distribution which falls with semi-major axis, given the strong preference of the General and Proper 2 models over the Proper 1 model, and further the preference of the General model over the Proper 2 model.

Although fitting models to the data in the visibility domain requires an order of magnitude longer to converge, we note that the overall result we present is robust to the methodology applied, and a visibility-domain analysis would conclude similarly. 
We verified this by producing from each of our three best-fit General, Proper~1 and Proper~2 models, visibility tables (gridded to the same uv-spatial frequencies as the data) from the Cycle~5, Apocentre-only pointing which contains the very highest SNR data \citep[see][]{Chittidi_rev}.
By subtracting these visibilities from the data in the visibility domain (after applying the {\tt CASA statwt} function to the data) we measured their relative $\chi^2$ values (directly from the visibilities) and imaged the residuals.
Via this method, the residuals strongly favor the General model, which leaves comparable residuals to those seen in Fig.~\ref{fig:dataresiduals} (i.e., no residual artifacts above $4\sigma$) and further, has the lowest $\chi^2$ by 350 (versus Proper~1) and 245 (versus Proper~2).

\subsection{Conclusion: an eccentric gradient in the Fomalhaut debris disk}
We find clear evidence that there is a gradient in the eccentricity distribution of Fomalhaut's planetesimals, and thus in its debris dust as observed with \textit{ALMA}. 
This finding suggests that the Fomalhaut debris disk is being shaped by the eccentric velocity divergence (EVD) effect, as discussed in \S\ref{sec:obsmodel}.
Most intriguing about the inferred power-law gradient value of the General model ($n_{\rm{pow}}{=}{-1.75}{\pm}0.16$) is its near-consistency with the classical expectation for secular interactions between an internal planetary perturber with distantly-separated outer planetesimals \citep[for which $n_{\rm{pow}}{=}{-1}$; see e.g.,][i.e. the same value that we fixed in the Proper 2 model, and fitted to the archival ALMA data]{Murray99}.
The value fitted to the newest data nevertheless appears steeper than this classical expectation (being discordant at the $4.6\sigma$ level, though we note this discrepancy disappears when we fit to the single configuration, lower-resolution mosaic of \citet{MacGregor17}, which is fully consistent with a power-law eccentricity gradient of $-1$).
We defer further discussion to the origin of this eccentricity profile to \S\ref{sec:rebound}.

\section{Analysis of the planetary system}
\label{sec:planets}
Since the inference of an eccentricity gradient in Fomalhaut's disk has important implications for the presence and orbital properties of an internal planet interacting with the disk, we next investigate planet properties plausible with this interpretation.
We first consider plausible single-planet scenarios that could be responsible for carving the disk structure in \S\ref{sec:planets}, i.e., based on the modelled properties of the disk inner edge and inner edge eccentricity.
Further, in \S\ref{sec:rebound} we consider if either such planetary scenario is preferred, and further, the plausible origins of the disk's eccentricity.
Specifically in \S\ref{sec:rebound_1} we conduct N-body modeling of planet--disk interactions consistent with the derived planet scenarios, and in \S\ref{sec:selfgrav} we analytically investigate whether disk self-gravity (in the presence of an internal planet perturber consistent with the derived planet scenarios) may be responsible for driving a steeper gradient (than $n_{\rm{pow}}{=}{-1}$) in to the disk.

\subsection{Planet properties consistent with \textit{ALMA} and \textit{JWST} observations} \label{sec:planProps}
On the assumption that a single, co-planar planet is responsible for carving the observed disk structure in the Fomalhaut debris disk (i.e., its inner--edge location, eccentricity, and eccentricity gradient), we consider two scenarios that can constrain such a planet's properties\footnote{For a discussion of the parameters one may derive for planets orbiting Fomalhaut at inclinations \textit{misaligned} with the disk, we refer the reader to \citet{Pearce21}.}.
Recent observations of Fomalhaut with the \textit{JWST NIRCam} instrument place constraints on planet masses of $M_{\rm{pl}}{\lesssim}200\,M_\oplus$ beyond $5''$ (38.3\,au) from the stellar center \citep{Ygouf+2024}.
Our analysis of single-planet scenarios are consistent with such a non-detection, though we find the plausible planet mass range well below this \textit{NIRCam} limit. 
Perhaps more important for constraining planetary properties are the \textit{JWST MIRI} observations \citep{Gaspar+2023}.
In these, the first evidence of an `intermediate belt' is presented, which has inner and outer edges of 83\,au and 104\,au respectively, with large corresponding eccentricities of 0.31 and 0.265.
We utilise these findings here, on the assumption that a planet could not reside within the region spanned by the intermediate belt, but instead, must be located either \textit{between} the \textit{ALMA}-observed `main belt' and the intermediate belt, or interior to the intermediate belt.
We present the single planet properties (derived in \S\ref{sec:pl1} and \S\ref{sec:pl2}) in Table~\ref{tab:planets}.

\subsubsection{A gap-carving planet?} \label{sec:pl1}
We first discuss the possibility that a single planet fully carved the gap up to the disk inner--edge directly. 
In this scenario, the \textit{ALMA} observations can place constraints on such a planet's semi--major axis and eccentricity, using equation 10 of \citet{Mustill12}, i.e., $(a_{\rm{inner}}-a_{\rm{pl}})/a_{\rm{pl}} {=}1.8\times(e_{\rm{inner}}\mu)^{1/5}$.
For a planet with a mass ranging from $M_{\rm{pl}}{\approx}1-16\,M_\oplus$ (or $\mu$ in the range $1.7\times10^{-6}$ to $2.5\times10^{-5}$ given the $2M_\odot$ mass of Fomalhaut, the mass range for which we justify below), this expression requires planet semi--major axes in the range $a_{\rm{pl}}{=}109-115$\,au, and eccentricities in the range $e_{\rm{pl}}{=}0.18-0.20$, with the innermost semi--major axis $a_{\rm{pl}}$ in this range corresponding to the highest values of $e_{\rm{pl}}$ and $M_{\rm{pl}}$.
We calculate these values based on the inner--edge disk semi--major axis (of $a_{\rm{inner}}{\sim}125\,$au) and a disk eccentricity at this location ($e_{\rm{inner}}{\approx}0.16$, estimated on an extrapolation from the disk center of the fitted values of $e$ and $n_{\rm{pow}}$). 

We justify the range of planet masses with the lower-limit corresponding to the timescale on which a given planet can plausibly secularly force its eccentricity into the disk, and at the upper-limit based on direct planetesimal clearing.
For example, applying equation~15 of \citet{Mustill09} with the range of derived $a_{\rm{pl}}$ and $e_{\rm{pl}}$ and known stellar mass, semi--major axes out to 150\,au require at least a $\mu\approx1.7\times10^{-6}$ body to be secularly forced within the age of the Fomalhaut system.
In addition, equation~10 of \citet{Mustill12} suggests that planets at 109\,au require masses of $\mu\approx2.5\times10^{-5}$ to force the inner-edge disk eccentricity.
Utilising \citet{Morrison15}, i.e. that planets clear material within $\Delta a=1.2\mu^{0.31}a_{\rm{pl}}$ of their semi-major axis, such a body would clear material inwards to 104\,au, i.e., the intermediate belt outer edge, thus we set a mass upper-limit based on this location.
We refer to this as the `Gap' planet in Table~\ref{tab:planets}.
We note that if the intermediate belt is not a long-lived planetesimal belt, planet mass constraints from \textit{JWST NIRCam} (i.e., $M_{\rm{pl}}{\lesssim}200\,M_\oplus$) would allow planet semi-major axes and eccentricities to span a slightly broader range of 100--120\,au and 0.18--0.23 respectively.

\begin{table}[]
    \centering
        \caption{Derived constraints for a single planet responsible either for directly carving the inner-edge gap of the main belt (`Gap') or clearing via a 2:1 resonance (`Resonant').}
    \begin{tabular}{l|ccc}
        \hline
        \hline
        Scenario & $a_{\rm{pl}}$ & $e_{\rm{pl}}$ & $M_{\rm{pl}}$ \\
        & [au] & -- & [$\mu$] \\ 
         \hline
        Gap & 109--120 & 0.20--0.23 & $1.5{\times}10^{-6}$ -- $2.5{\times}10^{-5}$ \\
        Resonant & 70--75 & 0.38--0.41 &
        $7.0{\times}10^{-6}$ -- $2.5{\times}10^{-5}$ \\
         \hline
    \end{tabular}
    \label{tab:planets}
\end{table}

The planet orbital parameters of this `Gap' planet all agree with those presented for Fomalhaut~b \citep{Kalas08}.
Although more recent observations and analysis of Fomalhaut~b imply this is no longer present \citep{GasparRieke2020, Ygouf+2024}, this analysis suggests that \textit{a} planet with properties consistent with Fomalhaut~b may be present in the system.

\subsubsection{A resonant-clearing planet?} \label{sec:pl2}
The presence of Fomalhaut's intermediate belt presents another plausible scenario in which the gap between the main belt and the intermediate belt was cleared by an internal planet's resonant interactions.
Such a planet would need to be internal to 83\,au based on the $e=0.31$ intermediate belt inner-edge, and have a strong resonance located between the inner edge of the main belt and the outer edge of the intermediate belt.
For a planet's (strongest) 2:1 mean motion resonance to reside within 2\,au from the mid-point of these belt locations, these must span semi-major axes of 70--75\,au. 
This semi-major axis range further requires a (narrower) range of planet masses of $7\times10^{-6}\lesssim \mu \lesssim 2.5\times10^{-5}$ for these planets to be responsible for the eccentricity at the inner-edge of the intermediate belt (i.e., we utilise once more $(a_{\rm{inner}}-a_{\rm{pl}})/a_{\rm{pl}} {=}1.8\times(e_{\rm{inner}}\mu)^{1/5}$, with $e_{\rm{inner}}=0.31$).
If these bodies also follow the power-law eccentricity gradient, then these would have eccentricities of 0.38--0.41.
We refer to this as the `Resonant' planet in Table~\ref{tab:planets}.

We note that in this resonant clearing scenario, if such a planet was responsible for driving the eccentricity distribution of the outer belt (which we measured as $e_f\propto a^{-1.7}$) then it could feasibly drive larger eccentricities in the intermediate belt given these would be much closer in their semi-major axes. 
By extrapolating the fitted eccentricity power-law to semi-major axes of 83--104\,au (i.e., those of the intermediate belt), we calculate eccentricities in the range 0.22--0.32, which overlap reasonably well with the values fitted to the inner and outer edges of the \textit{JWST} observations of 0.31 and 0.265 respectively \citep[][]{Gaspar+2023}.
Whilst not stated explicitly in the work of \citet{Gaspar+2023}, their \textit{JWST} analysis corroborates the finding that Fomalhaut hosts an eccentricity distribution that falls with semi-major axis (given the decrease in eccentricity from the intermediate belt inner to outer edge).

\begin{figure*}
    \centering
    \includegraphics[width=1.0\textwidth]{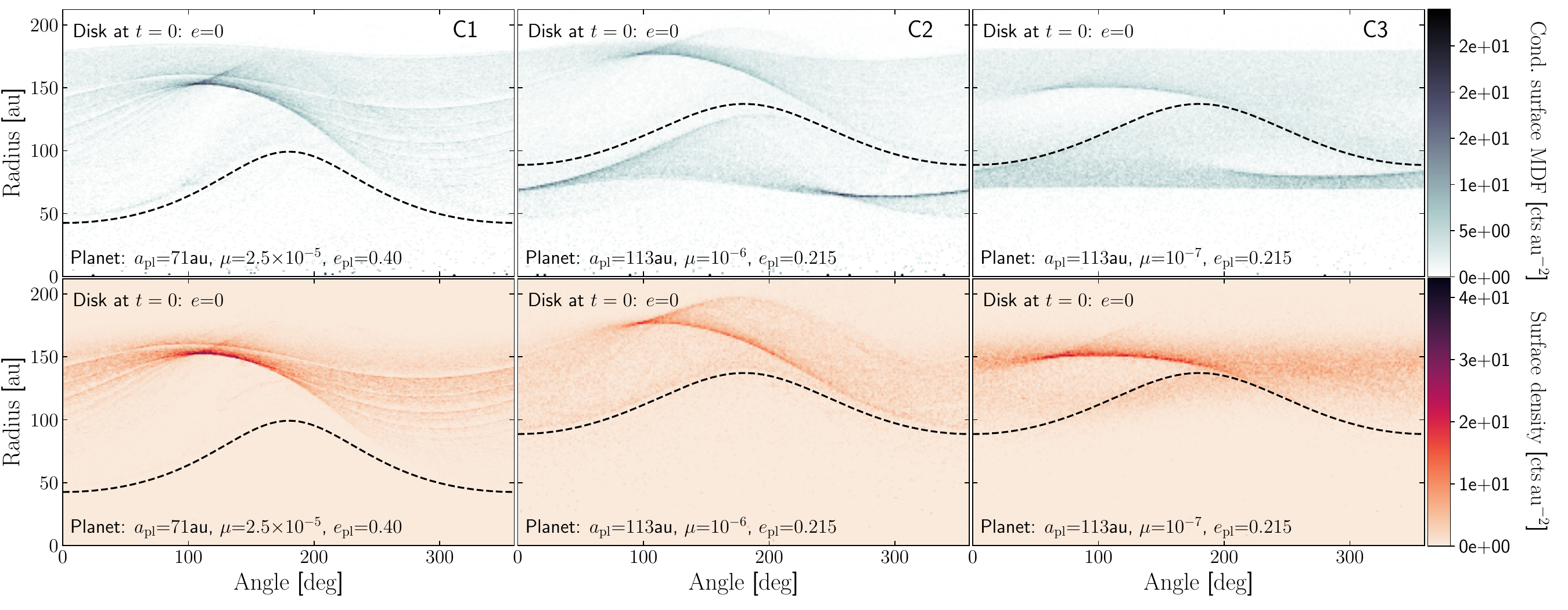}
    \includegraphics[width=1.0\textwidth]{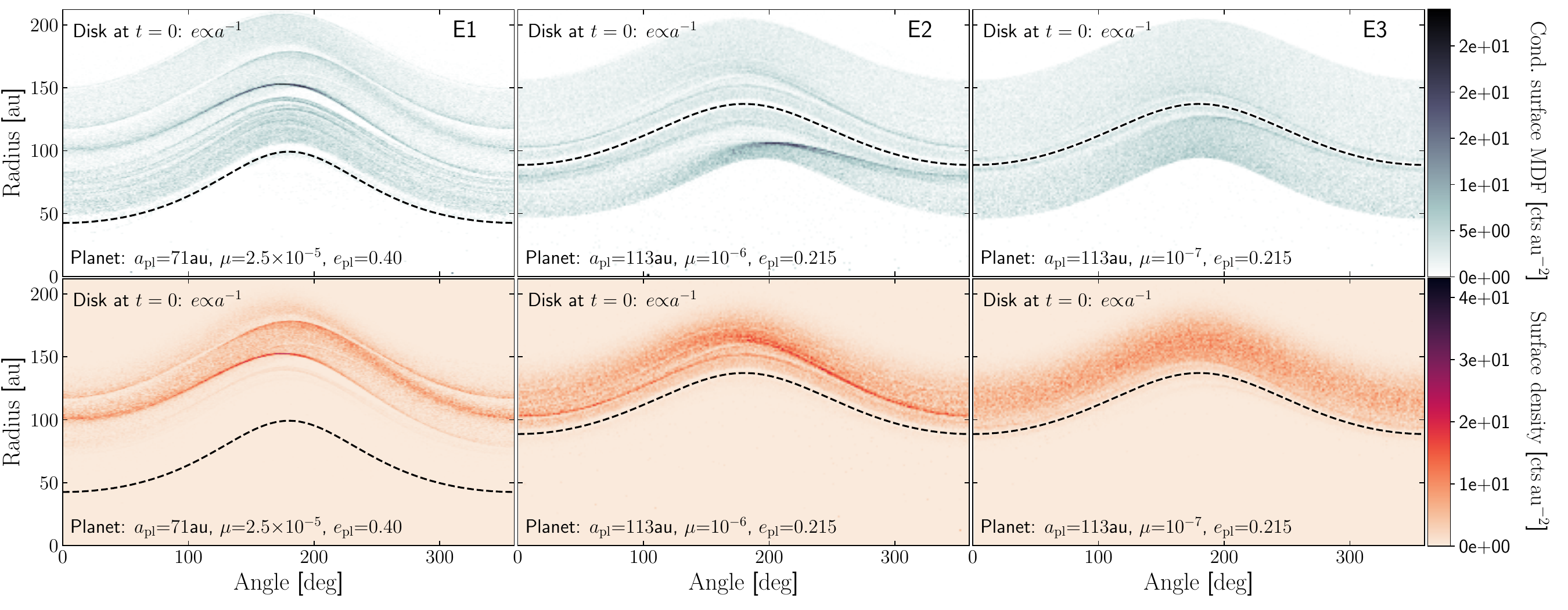}
    \caption{Plots for the surface density maps from the {\tt REBOUND} simulations in $r-\phi$ space. Upper two rows: (conditional surface mass density function (MDF), and absolute surface density) simulations with initially circular disk conditions; lower two rows: simulation outcomes with initially eccentric disk conditions presented likewise.
    Specific planet and disk assumptions are presented on each plot, with black-dash lines showing the planet orbit.
    Initially circular disk conditions do not yield stable disks at 440\,Myr, whereas scenarios which are initially eccentric yield stable eccentric disks at 440\,Myr.
    The same information is tabulated in Table~\ref{tab:reboundscenarios}.}
    \label{fig:rebound}
\end{figure*}

\section{The origin of Fomalhaut's eccentric disk} \label{sec:rebound}
\subsection{N-body modelling of the Fomalhaut debris disk} 
\label{sec:rebound_1}
In this section we investigate the plausible origins of the disk eccentricity, by considering the stability of the disk over the full 440\,Myr age of the star \citep{Mamajek2012}.
We base this section on the outcome of our earlier modelling, and the plausible (single) planet scenarios consistent with observations.
In all cases, we consider a scenario with a single star (Fomalhaut~A) orbited by a single (gravitating) eccentric planet, of substantially lower mass.
To determine the stability of the Fomalhaut debris disk, we consider the interaction of this star-planet system with a population of test particles (the effects of the disk self-gravity is explored in the Section \ref{sec:selfgrav}), which sample the hypothetical birth orbital element distribution of the Fomalhaut main and intermediate belts.

\begin{table*}[t]
    \centering
    \caption{Here we show all six {\tt REBOUND} scenarios simulated in this work, starting with disk properties (Disk extent and eccentricity) and planet properties (planet--stellar mass ratio and semi--major axis) and describe the stability of the disk and comment on any observed features present. }
    \begin{tabular}{l|cc|ccc|l}
        \hline
        \hline
         & Disk semi--&Disk & Planet &Planet semi--&Planet&   \\
        ID& major axis & eccentricity & mass ratio & major axis & eccentricity & Comments\\
        & extent [au] & profile & & [au] & &  \\

        \hline
         C1 & 70--180 & $e=0$ & $2.5 \times 10^{-5}$ & 70.87 & 0.4 & Disk disrupted \\
         C2 & 70--180 & $e=0$ & $1.0 \times 10^{-6}$ & 112.9 & 0.215 & Disk disrupted \\
         C3 & 70--180 & $e=0$ & $1.0 \times 10^{-7}$ & 112.9 & 0.215 & Disk disrupted \\
        \hline
         E1 & 70--180 & $e\propto a^{-1}$ & $2.5 \times 10^{-5}$ & 70.87 & 0.4 & Gap carved, gap in main belt, spirals \\
         E2 & 70--180 & $e\propto a^{-1}$ & $1.0 \times 10^{-6}$ & 112.9 & 0.215 & Gap carved, weak spirals  \\
         E3 & 70--180 & $e\propto a^{-1}$ & $1.0 \times 10^{-7}$ & 112.9 & 0.215  & Gap carved, but very shallow  \\
        \hline
    \end{tabular}
    \label{tab:reboundscenarios}
\end{table*}

\subsubsection{{\tt REBOUND} setup}
To investigate the stability of orbits in the Fomalhaut system, we use {\tt REBOUND} \citep{Rein+2012}.
We adopt the {\tt WHFAST} symplectic integrator \citep{Rein+2015} with a fixed timestep set to 1 per cent of the initially innermost particles orbital period\footnote{Such a timestep choice is consistent with {\tt REBOUND} simulations of debris disks in the literature, e.g., \citet{Dong2020} and \citet{Pearce2024}, and consistent with the advice provided in the {\tt REBOUND} API Documentation; see comment on `Timestepping' in \url{https://rebound.readthedocs.io/en/latest/simulationvariables/}.}, using democratic heliocentric coordinates \citep[i.e., where positions are measured relative to the star, and the corresponding canonical momenta are measured relative to the barycentre;][]{ReinTamayo2019}. 
In total, we consider six scenarios (that investigate three planet setups, and two disk setups) that are summarised in Table \ref{tab:reboundscenarios} and described below.

We explore both the `Gap' and `Resonant' planet scenarios (as per Table~\ref{tab:planets}) for either initially circular or initially eccentric test particle disks.
For the Resonant planet we consider a single mass of $M_{\rm pl} = 2.5 \times 10^{-5} M_{*}$, a semi-major axis of $a_{\rm pl} = 70.87\,$au and an eccentricity $e_{\rm pl} = 0.4$. Scenarios with a `1' in their ID include this planet.
For the Gap planet, we consider two masses of $M_{\rm pl} = 1.0\times10^{-6} M_{*}$ and $M_{\rm pl} = 1.0\times10^{-7} M_{*}$, both with a semi-major axis of $a_{\rm pl} = 112.9\,$au, and eccentricity $e_{\rm pl} = 0.215$. Scenarios with either a `2' or `3' in their ID include this planet.
Adopting units in which $G M_{*} = 1$ and unit length equal to the planets semi-major axis, each planet has an orbital period $\approx 2 \pi$ (strictly the orbital period is slightly shorter due to the finite planet mass). 

We consider two initial scenarios, either with all particles initially circular, or initially eccentric.
In the case of the initially circular test particles, we distribute these with $e=0$. 
Scenarios with a `C' in their ID relate to this initial disk setup.
In the case of the initially eccentric test particles, we distribute these with $e = e_{\rm pl} (a/a_{\rm pl})^{-1}$ (i.e., consistent with our modelling).
Scenarios with a `E' in their ID relate to this initial disk setup.
We setup the disk with $2 \times 10^5$ test particles uniformly distributed between $70 (a_{\rm pl}/\mathrm{au})^{-1}$ and $180 (a_{\rm pl}/\mathrm{au})^{-1}$.
We note that this broad radial distribution spans all radii where Fomalhaut's main belt and intermediate belt have been observed. 
In all cases, we randomly distribute the particles uniformly in semi-major axis-mean anomaly space. 
To avoid issues with the planetary point mass singularity we adopt {\tt REBOUND}'s gravitational softening with a gravitational softening parameter set to $b = 10^{-6} [a_{\rm pl}]$. 
This choice of softening radius is well within the planets hill sphere for all scenarios considered.

\subsubsection{{\tt REBOUND} results}
We run simulations E1, E2, E3, C2 and C3 for the full 440\,Myr lifetime of Fomalhaut. 
We run simulation C1 for just 1\,Myr as by this time the disk is already disrupted (in the sense that large sections of the disk have been broken up or cleared by interaction with the planet). 
We show the outcomes of these six scenarios in Fig.~\ref{fig:rebound}, firstly in terms of their `conditional' mass surface density distributions (i.e., the simulation outcomes), and secondly to aid interpretation, as their surface density distributions convolved with a specified radial profile.
These convolved distributions convolve the results of the {\tt REBOUND} simulations, with a mass per unit semi-major axis of
\begin{equation}
 M_a = \frac{\sqrt{2 \pi} a}{w_r} \exp \left( \frac{-(a - a_0)^2}{2 w_r^2} \right),
\end{equation}
consistent with the modeling in Section \ref{sec:obs}. 
Presenting the data in this manner comes with important caveats.
Firstly, this profile choice effectively masks all emission not imaged by \textit{ALMA}. 
Thus, whilst the simulations can inform us of the stability of orbits in Fomalhaut's intermediate belt, we restrict our investigation here purely to Fomalhaut's main belt.
By extension, this further downplays the importance of the planet's gap-carving since this hides the fact that lower mass planets are unable to carve deep gaps in the particle distributions (one should thus consider both plots per scenario).
Secondly, due to our ignorance about the initial conditions of the Fomalhaut disk, the chosen mass per orbit is, by necessity, arbitrary. 
The mass per orbit is not affected by secular perturbations so, away from resonances and close encounters, the planet has no way to modify the mass on orbits. 
Overall, the {\tt REBOUND} simulations are thus not expected to reproduce either the \textit{ALMA} and \textit{JWST} observations, but instead inform us about the stability of the orbits within the belt and identify features (e.g., gaps or spiral density waves) that may be expected to be observed for given planetary properties. 

With these caveats in mind, we find (in all cases) that scenarios initialised with circular disks do not produce stable eccentric belts.
In the three cases C1, C2 and C3, we find that these all end their simulations with disrupted particle distributions (inconsistent with observations).  
In contrast, we find that scenarios initialised with eccentric disks (E1, E2 and E3) survive the full 440\,Myr simulation timescale, and result in eccentric disks consistent with those observed for Fomalhaut.
In addition, both the Resonant and Gap planet scenarios carve gaps in the location between the main and intermediate belts (in the lowest mass planet case, this gap is relatively shallow in comparison to the more massive planet scenarios). 
One intriguing feature in the E1 scenario is the formation of a spiral density wave (due to the massive Resonant planet's interaction with the disk).
Whilst no such features have been observed in the Fomalhaut disk as yet, deeper, higher resolution observations could be used to investigate the plausibility of this scenario and constrain the mass of a planet with these orbital parameters.

Overall, the {\tt REBOUND} simulations lend weight to two important conclusions.
Firstly, the simulations suggest that Fomalhaut's disk may need to have been \textit{formed} as an eccentric disk in order to survive $\sim440\,$Myr evolution (i.e., formation within the protoplanetary disk).
Secondly, and by extension, whilst a single-planet could be responsible for carving the inner-edge of the Fomalhaut main belt (or the gap between the intermediate and main belts) this same planet may not be capable of forcing its own eccentricity into an initially circular planetesimal belt via secular planet--disk interactions to yield an eccentric disk with Fomalhaut's properties.
In the case of the Gap planet, this body can plausibly clear material from the main belt's inner edge (and the intermediate belt's outer edge) if this is situated in an initially eccentric debris belt, i.e., via close planet--planetesimal encounters.
In the case of the Resonant planet, this could plausibly clear material in the same location due to the overlap with its 2:1 mean motion resonance and subsequent planetesimal ejections. 

\subsubsection{Comparison with eccentric protoplanetary disks}
The conclusion that eccentric disks that survive to the stellar main sequence may need to be produced during the protoplanetary disk phase is consistent with the conclusion of \citet[][]{Kennedy20} who argued this from the independent standpoint of the apparent narrowness of the debris disks of Fomalhaut and HD~202628.
There are indeed a (small) number of known eccentric protoplanetary disks, such as MWC~758, HD~100546 and IRS~48 \citep[][all of which are hosted by ${\sim}2\,M_\odot$ stars]{Dong+2018, Fedele+2021, Yang+2023}, as well as the eccentric circumbinary disk of IRAS~04158+2805 \citep[which is hosted by a lower--mass mid--M SpT binary;][]{Andrews+08, Ragusa+2021}. 
The eccentric debris disks of HD~202628 and HD~53143 are hosted by ${\sim}$Solar--mass stars \citep{Faramaz19, MacGregor22}, and those of HD~38206 and HR~4796 likewise orbit ${\sim}2\,M_\odot$ (A--type) stars \citep{Olofsson2019, Booth+2021}. 
Whether this bias towards earlier-type, intermediate-mass stars is physical or observational should be determined in future work.

\textit{A plausible origin of Fomalhaut's eccentric disk?}
The Fomalhaut debris disk's eccentricity could have resulted from the formation of a planet (with properties consistent with those in Table~\ref{tab:planets}) within the protoplanetary disk, whereby initial planet--gas disk interactions determined the eccentricity profile of the disk, and the eccentricity of the planet. 
In such a scenario, an eccentricity profile which decreases with radius, is expected \citep[e.g.][]{Teyssandier+2016}.
This finding follows from the fact that eccentricity profiles that are either flat or increasing with radius tend to differentially precess as a result of gas pressure forces.
In the case of Fomalhaut, following the dissipation of its protoplanetary disk gas (and any remnant primordial dust), subsequent planet--disk interactions with the eccentric planetesimal belt plausibly sculpted its disk inner and outer edges, which over the course of 440\,Myr resulted in the belt morphology, whilst maintaining it's initial eccentricity distribution (or one very similar).

\subsection{Additional physics? The role of disk self-gravity} \label{sec:selfgrav}

We have thus far explored the possible origin of Fomalhaut's disk eccentricity assuming a stationary planet perturbing a \textit{massless} debris disk. However, additional physical effects, such as the  influence of a massive, self-gravitating debris disk, may affect the outcomes.

As mentioned in Section \ref{sec:obsmodel},
a planet within a debris disk induces (secular) forced planetesimal eccentricities such that $e_f(a) \propto a^{-1}$\citep[][]{Murray99}.
This, however, applies specifically to a massless debris disk ($M_d =0$), i.e., composed of test-particles. 
By contrast, if the disk is massive, its self-gravity can steepen the forced eccentricity profile \citep{Sefilian2024}: namely, to as much as $e_f(a) \propto a^{-4.5}$, depending on $a_p/a_{\rm in}$ and the mass distribution within the disk \citep[see also,][]{Sefilian+2021}.
This occurs when the disk-induced apsidal precession rates of the planet and/or the planetesimals -- \textit{absent in massless disk models} -- dominate over the planet-induced precession of planetesimal orbits.
For $a_p \sim a_{\rm in}$, this condition  generally translates to $M_d/m_p \gtrsim 1$, and would require $M_d/m_p \lesssim 1$ for $a_p \lesssim a_{\rm in}$. 
Interestingly, under the same conditions on $M_d/m_p$, a similar effect arises in  the disk's vertical aspect ratio $h$ if the planet--disk system is misaligned: rather than remaining constant with radius (as expected for $M_d = 0$), $h$ may instead decrease with distance from the star if $M_d \neq 0$ \citep{Sefilian2025}. This effect, however, is not accounted for in this study. 

Without presuming specific planet masses, \textit{ALMA} observations imply a total disk mass (i.e., including the largest, colliding planetesimals) consistent with the range inferred by the collisional modelling of \citet{Krivov21}, which estimates $M_d = 1.8{-}360\,M_\oplus$.
If true, disk gravity could potentially explain the steep  $e_f(a)$ profile indicated by our modelling. While it may be tempting to test this hypothesis using the results of \citet{Sefilian2024}, their study is a proof-of-concept based on a semi-analytical framework that accounts for the axisymmetric component of the disk potential, but ignores the non-axisymmetric counterpart. In principle, this limitation can be addressed using 
existing orbit-averaged, secular codes (e.g., the linear, \texttt{N-RING} code of \citet{Sefilian+2023} or the more accurate \texttt{GAUSS} code of \citet{Touma2009}) and/or direct $N$-body codes; however, this is more suited to a separate study (Sefilian et al., in prep.).

\section{Conclusions} \label{sec:conclusions}
We present a new model of Fomalhaut's eccentric debris disk by fitting parametric debris disk models to archival \textit{ALMA} observations (with a synthesised physical resolution 4--6\,au).
The models are developed from those introduced by \citet{LynchLovell21} and \citet{LovellLynch2023} whereby a gradient in the forced eccentricity parameter, $e(a)\propto a^{n_{\rm{pow}}}$, is introduced to fit the known disk asymmetries; a brighter apocenter versus pericenter, and a broader pericenter versus apocenter.
We collectively term this physical effect within disks `Eccentric Velocity Divergence'.
The best-fit parameter determinations from the modeling find parameters broadly consistent with those from other models, but with the inclusion of a steep, negative eccentricity gradient as per $e\propto a^{-1.75{\pm}0.16}$.
By analysing the goodness-of-fit, we deem that such a description provides a novel interpretation of this debris disk, and is statistically preferred to models with constant free and forced eccentricity distributions through the disk.
To our knowledge, this is the first reported eccentricity gradient in a debris disk.
The value we derive is very close to that expected from classical expectations of (massless) disk--planet interactions $e_f \propto a^{-1}$, suggesting the gradient may be forced by a planetary perturber.

Based on the \textit{ALMA} modeling, and published data from \textit{JWST NIRCam} and \textit{MIRI}, we quantify planetary orbital and mass constraints for two co-planar single-planet scenarios consistent with observations.
One scenario describes a 109--115\,au planet that has directly cleared material up to the inner--edge of Fomalhaut's `main belt' (as imaged by \textit{ALMA}).
The second posited scenario describes a 70--75\,au planet that has cleared the disk inner--edge at its 2:1 mean motion resonance, with this planet sitting interior to the \textit{JWST}-imaged `intermediate belt'.
In both cases, the implied planet mass and semi-major axis ranges are below sensitivity thresholds for existing planet detection methods.

We assess the stability of the Fomalhaut debris disk by conducting N-body {\tt REBOUND} modeling.
We set up six sets of initial conditions, based on three planets, and two initial conditions for the disk, with this either being circular or eccentric.
We show that only scenarios with an initially circular disk are stable/finalise as eccentric disks after the 440\,Myr simulation lifetime (matching that of the age of Fomalhaut).
These findings may suggest that planet-disk interactions are primarily responsible for sculpting the disk's morphology (i.e., its inner-edges, and as-per the \textit{JWST} observations, gaps in the disk), but not its eccentricity, and thus, that Fomalhaut's eccentric ring was plausibly born eccentric.
We discuss caveats with these models (such as the exclusion of disk self-gravity) and propose further simulations to investigate the robustness of these findings for massive debris disks.

Finally, we highlight that the code used to model Fomalhaut's surface density distribution in this work (equations~\ref{eq:1}--\ref{eq:4}) is available publicly on \url{ https://github.com/astroJLovell/eccentricDiskModels}. 
We release this to enable our work to be reproduced easily, and such that others can utilize this to model the surface density profiles of other optically thin disks.

\section{Software and third party data repository citations} \label{sec:data}
This paper makes use of the following \textit{ALMA} data: ADS/JAO.ALMA 2013.1.00486.S, ADS/JAO.ALMA 2015.1.00966.S and ADS/JAO.ALMA 2017.1.01043.S. ALMA is a partnership of ESO (representing its member states), NSF (USA) and NINS (Japan), together with NRC (Canada), MOST and ASIAA (Taiwan), and KASI (Republic of Korea), in cooperation with the Republic of Chile. The Joint \textit{ALMA} Observatory is operated by ESO, AUI/NRAO and NAOJ. 
The National Radio Astronomy Observatory is a facility of the National Science Foundation operated under cooperative agreement by Associated Universities, Inc.

\facilities{Atacama Large Millimeter/subMillimeter Array (ALMA)}

\software{This research made use of the following software packages:
          {\tt CASA} \citep{CASA07}
          {\tt DS9} \citep{ds9}.
          {\tt emcee} \citep{FM13};
          {\tt RADMC-3D} \citep{Dullemond12};
          {\tt REBOUND} \citep{Rein+2012}.
}

\vspace{-3mm}
\section*{Acknowledgments}
We thank the anonymous referee for their comments and suggestions which greatly improved the clarity of our investigation.
J.B.L. acknowledges the Smithsonian Institute for funding via a Submillimeter Array (SMA) Fellowship, and the North
American ALMA Science Center (NAASC) for funding via an ALMA Ambassadorship. J.B.L. dedicates this study to the memory of Henry Chesters, with whom he discussed this with great interest in their final conversation, and for HC's life-long encouragement of his scientific endeavors. 
A.A.S. is supported by the Heising-Simons Foundation through a 51~Pegasi~b Postdoctoral Fellowship.
We acknowledge the operational staff and scientists involved in the collection of data presented here.

\appendix{}
\section{Data masking}
\label{appX}
\begin{figure}
    \centering
    \includegraphics[width=1.0\columnwidth, trim={0cm 0.0cm 0cm 0cm},clip]{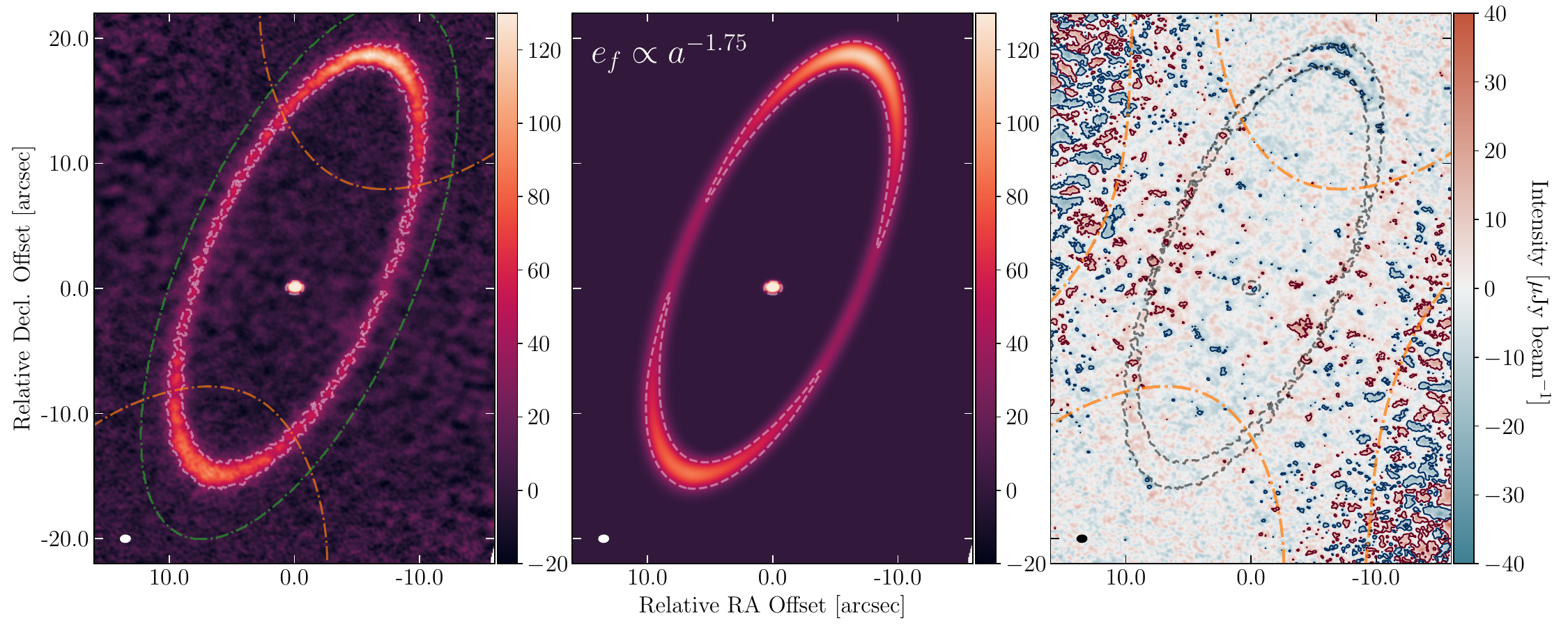}
        \caption{
        Left: \textit{ALMA} data. Center: Best--fit $e_f(a)\propto a^{-1.75}$ model (on same image scale). Right: residual emission after best-fit model subtraction. Contours are shown at the $\pm3\sigma$ level in the residuals, and at the $5\sigma$ level for the data/model. Beams are shown in the lower--left of each plot.
        In the left and right plots in amber dash-dot we overplot the primary beam response at the 0.66 level (and in addition at the 0.4 level on the right plot). We show in the left plot in green dash-dot the 200\,au spatial mask discussed in \ref{appX}.}
        
    \label{fig:mask}
\end{figure}

We present on Fig.~\ref{fig:mask} the result of fitting the General model while only masking pixels within a 200\,au (projected) distance from the star, that is, without masking by the primary beam.
We show the spatial mask on the left panel of this figure (in green dash-dot), and also show the primary beam response at the 0.66 level (in orange dash-dot) which defines the inner boundary of any fits that incorporate a primary-beam mask. The regions of the two ansae between the green and orange dash-dot lines define the mask used in the main body of this work.
It is evident that along the disk minor axis, the primary beam response is significantly reduced in comparison to the disk ansae, and whilst the minor axes appear to be slightly better fit in comparison to the General model with a primary beam mask, the fit is significantly worse at the apocentre justifying the choice to fit this model only at the two disk ansae.

\section{Model verification} \label{appB}
\begin{figure*}
    \centering
    \includegraphics[width=1.0\textwidth]{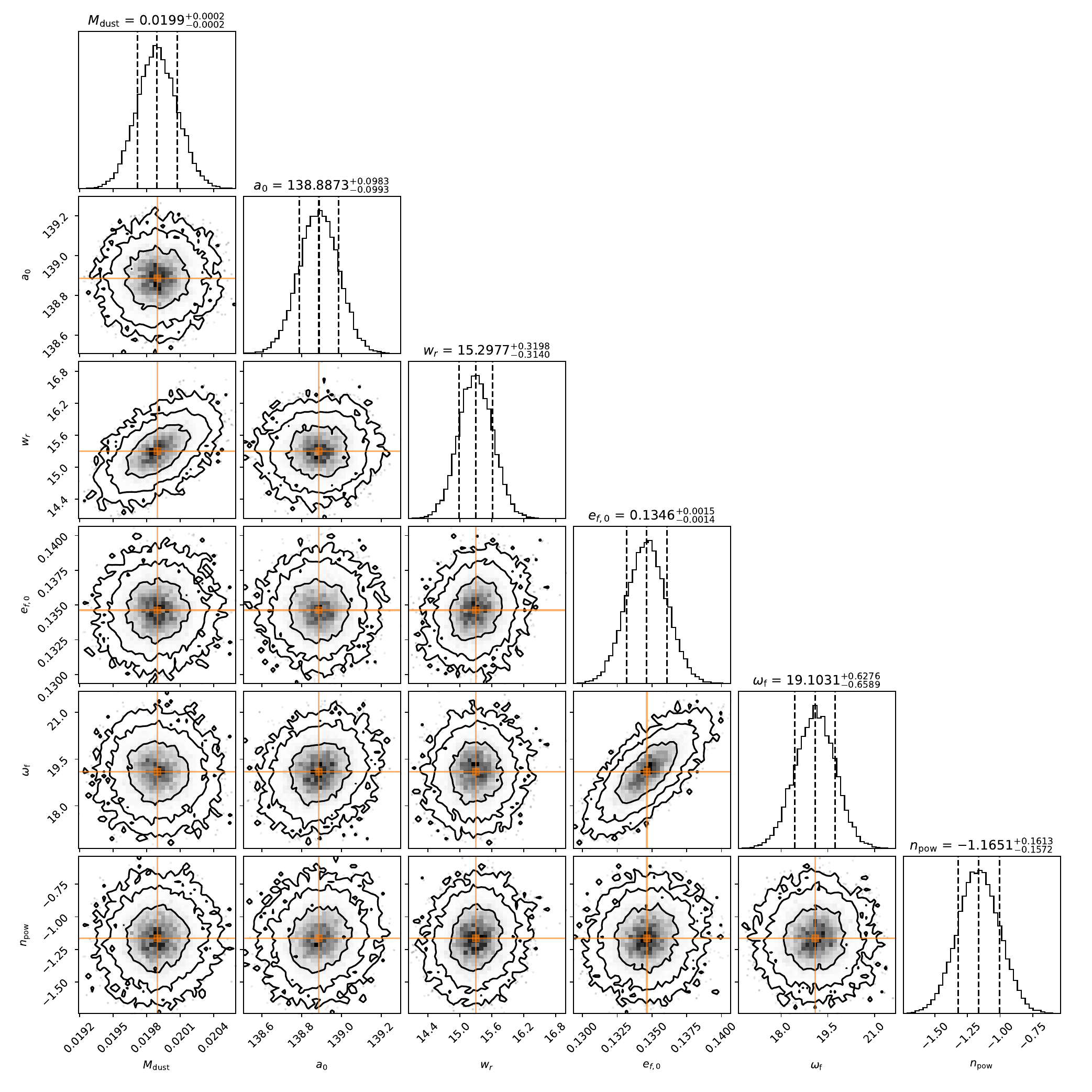}
    \caption{Corner plots for the {\tt emcee} chains (minus 1000 steps covering burn-in) showing the 6 parameters fitted to the low--resolution data of \citet{MacGregor17} and \citet{Kennedy20}, i.e. model `Verify'. We find comparable parameter uncertainties to those presented in \citet{Kennedy20} and a lower value of the rms in the residual map than the `uniform' model presented by \citet{Kennedy20}. }
    \label{fig:cornersverify}
\end{figure*}

We verify the modeling and fitting methodology by fitting a 6--parameter model to the archival \textit{ALMA} cycle--3 \textit{only} data for Fomalhaut, first presented in \citet{MacGregor17} and subsequently in \citet{Kennedy20}.
We refer to this low-resolution model as `Verify'.
We find that the parameter distributions and mean values of the fitting are consistent with those previous works \citep[i.e.,][]{MacGregor17, Kennedy20}, albeit with a slightly larger (mean) forced eccentricity (which is most likely due to the difference in eccentricity profile we fit here). 
Such consistency provides assurance that the fitting methodology we conduct here provides reasonable uncertainties on the fitted parameters.
We present in Fig.~\ref{fig:cornersverify} the posterior distribution after removing burn--in chains, the best-fit values on Table~\ref{tab:results}, and the best-fit model and residuals in Fig.~\ref{fig:residsBonus}.
Most interesting, is that we find a negative eccentricity gradient is likewise required with this model to fit even the lower-resolution data, with $n_{\rm{pow}}{=}-1.16{\pm}0.16$, and that this model setup can account for the disk width and brightness asymmetries \citep[presented by][]{Kennedy20}.
One improvement is that the  model we present here fully accounts for the north-west positive emission that is still present in the apocenter residuals of \citet{Kennedy20}.
In verifying this fitting methodology, we thus find that evidence of an eccentricity gradient in Fomalhaut's disk existed prior to its observations at higher--resolution.

\section{Residual emission maps and {\tt emcee} posterior distributions} \label{appA}
\begin{figure*}
    \centering
    \includegraphics[width=1.0\textwidth, trim={0cm 0.0cm 0cm 0cm},clip]{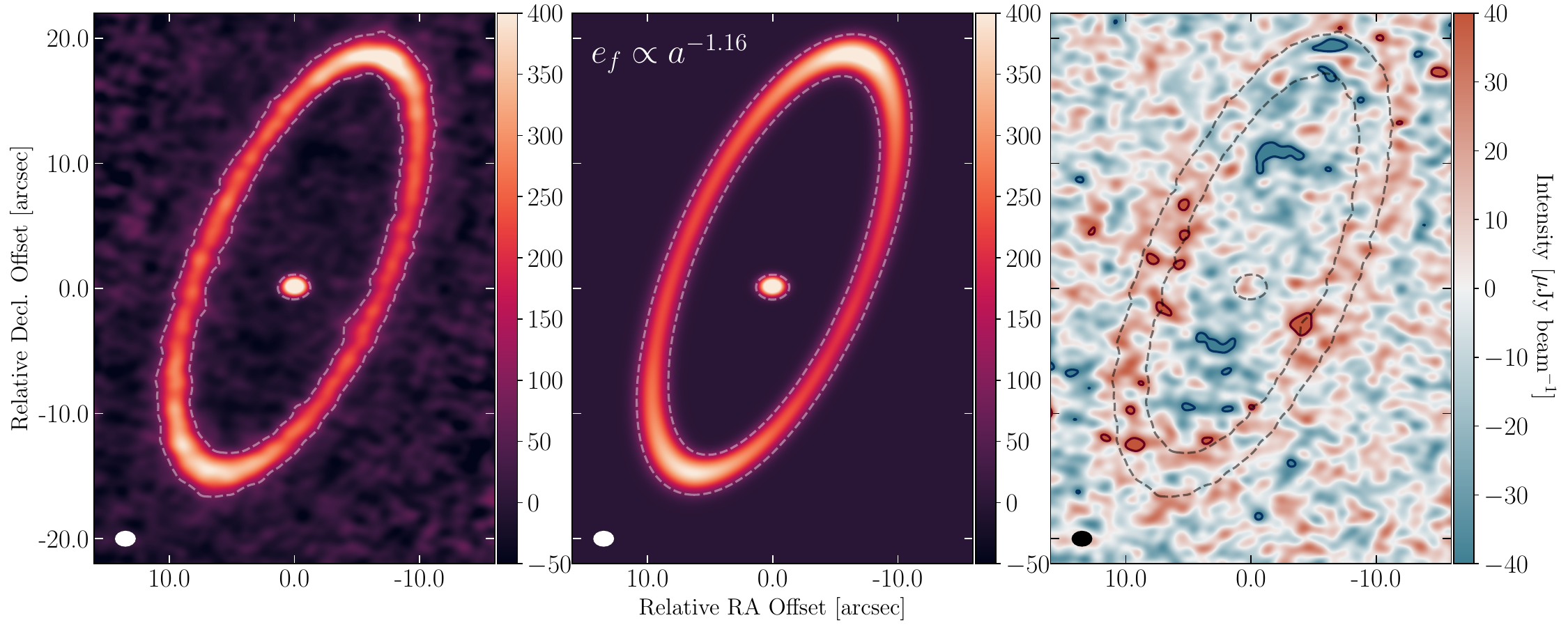}
        \caption{Left: \textit{ALMA} data as presented in \citet{MacGregor17} and \citet{Kennedy20}. Center: Best--fit model for this data. Right: residual emission (data minus best-fit model). 
    Contours are shown at the $\pm3\sigma$ level in the residuals, and at the $5\sigma$ level for the data/model. Beams are shown in the lower--left of each plot. In all plots, north is up, east is left. The strong peak at ($-4''$,$-3''$) is the source `GDC' as noted in both \citet{Gaspar+2023} and \citet{KennedyLovell2023}.}
    \label{fig:residsBonus}
\end{figure*}

\begin{figure*}
    \centering
    \includegraphics[width=1.0\textwidth]{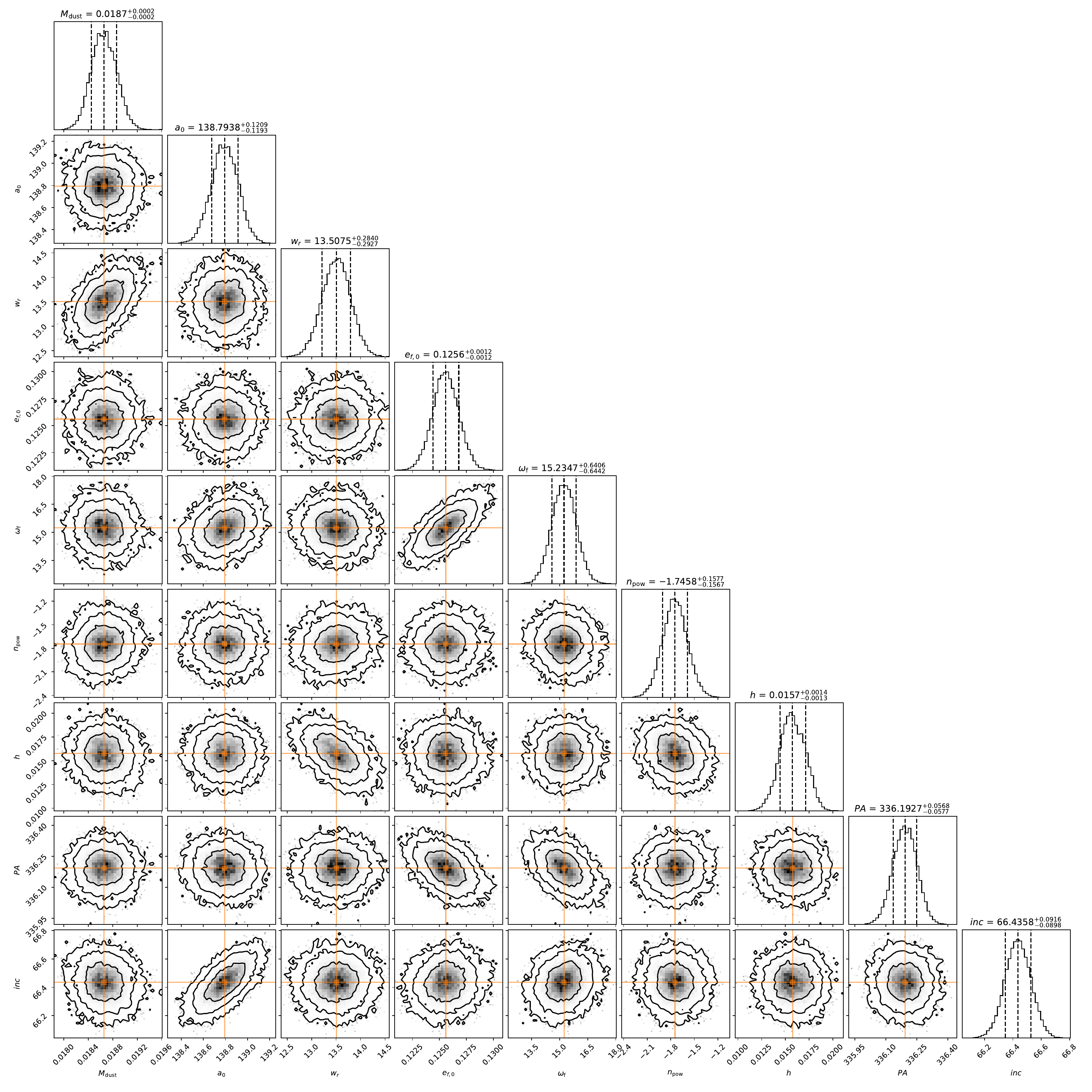}
    \caption{Corner plots for the {\tt emcee} chains (minus burn-in steps) showing the 9 parameters fitted in this study for the General model. All show the well-behaved features of  Gaussian distributions, that are either circular, or with little degeneracy (e.g., between $f_{\rm{M,\,dust}}$ and $w_r$, and between $e$ and $\omega_f$).}
    \label{fig:corners}
\end{figure*}

\begin{figure*}
    \centering
    \includegraphics[width=1.0\textwidth]{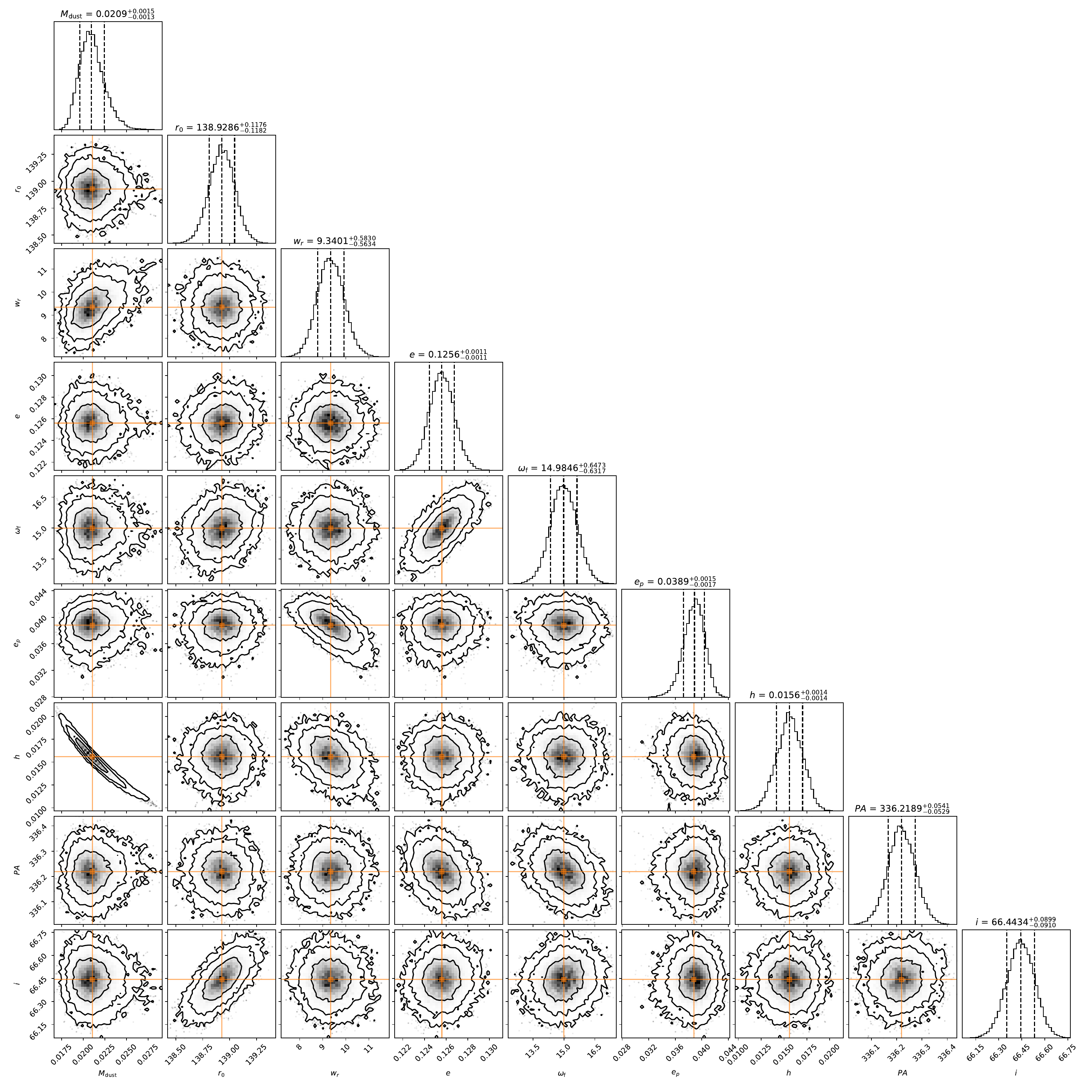}
    \caption{Corner plots for the {\tt emcee} chains (minus burn-in steps) showing the 9 parameters fitted for the Proper 1 model. All show the well-behaved features of  Gaussian distributions, that are either circular, or with little degeneracy (e.g., between $f_{\rm{M,\,dust}}$ and $w_r$, and between $e$ and $\omega_f$).}
    \label{fig:corners_ep1}
\end{figure*}

\begin{figure*}
    \centering
    \includegraphics[width=1.0\textwidth]{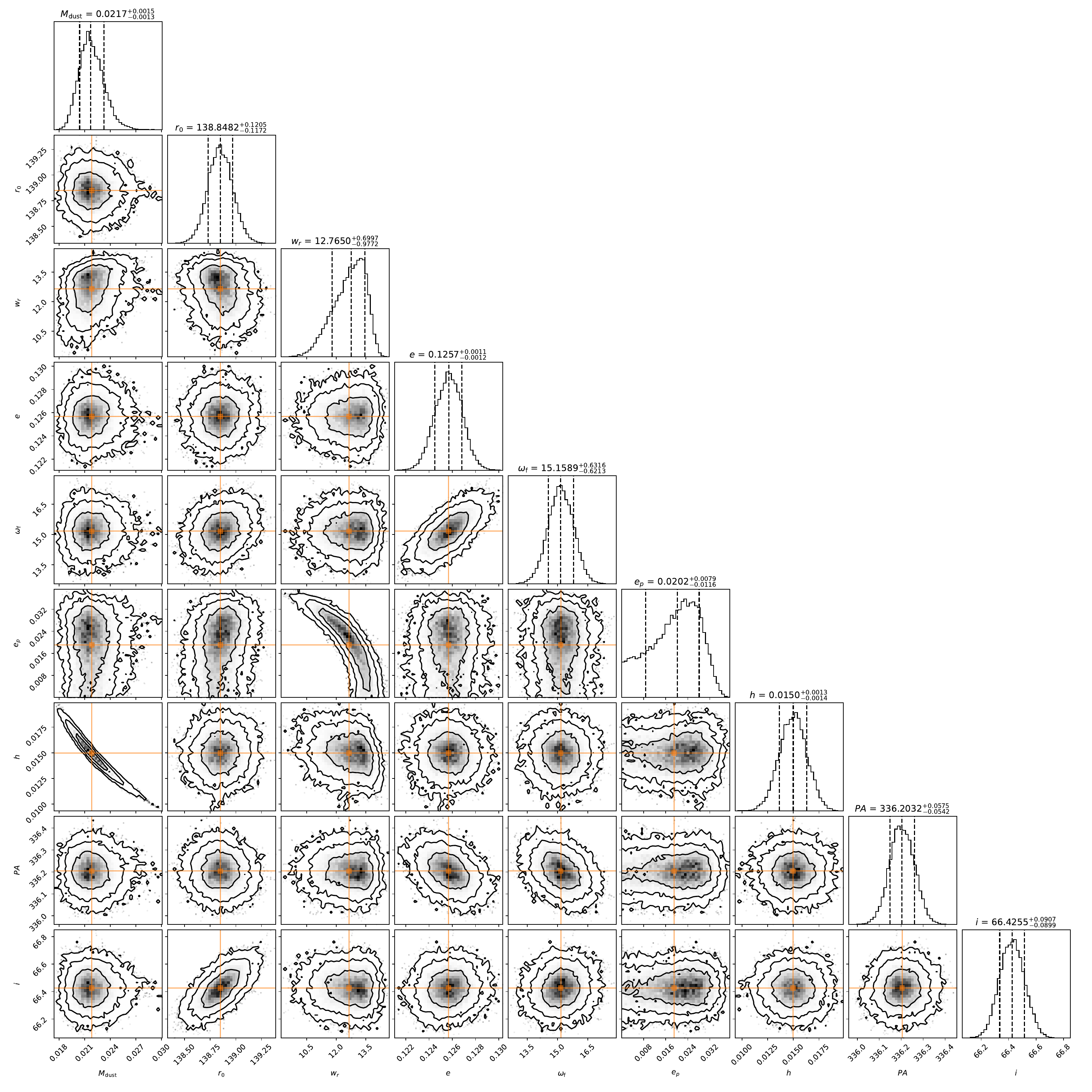}
    \caption{Corner plots for the {\tt emcee} chains (minus burn-in steps) showing the 9 parameters fitted for the Proper 2 model. All show the well-behaved features of  Gaussian distributions, that are either circular, or with little degeneracy (e.g., between $f_{\rm{M,\,dust}}$ and $w_r$, and between $e$ and $\omega_f$).}
    \label{fig:corners_ep2}
\end{figure*}

Here we present for the 6--parameter (low--resolution) `Verify' model, the corner plots in \ref{fig:cornersverify} and the data, model and residual maps in Fig~\ref{fig:residsBonus}.
We present the corner plots for the General, Proper 1 and Proper 2 models in Figs.~\ref{fig:corners},~\ref{fig:corners_ep1} and ~\ref{fig:corners_ep2}, and the data, model and residual maps for the Proper 1 and Proper 2 models in Fig.~\ref{fig:residseFree}.
In all cases, the first 1000 steps were removed from the {\tt emcee} chains (far in excess of the number of burn-in steps). 

\section{Model convergence} \label{appC}
We find by fitting initial rounds of {\tt emcee} (for the 9 free parameter General disk model) that the burn-in stage is approximately a few hundred steps, and that the auto-correlation time ($\tau$) is $\approx120$ steps. 
For the General and Proper 1 models, we measure the auto-correlation function ($\tau$) versus step length for all chains, and find this value asymptotes (with no deviation by more than 1\%) to a parameter-mean $\tau$ of $\approx120$.
In the case of the Proper 2 model, we measure a  mean $\tau$ of $\approx150$, and for the Verify model a mean $\tau$ of $\approx60$.
As stated in the {\tt emcee} guidance \citep{FM13}, convergence is achieved when the number of steps exceeds ${>}50\times \tau$. 
As such, we run the Verify, General, Proper 1 and Proper 2 models for $3000$, $6000$, $6000$ and $7500$ steps respectively and verify that convergence is indeed met in all cases. 
In all cases, we find after 300--500 steps the chains are burned in.
To derive the best-fit posterior distributions presented in Table~\ref{tab:results}, we removed the first half of the total steps from all chains.
We used 40 walkers in all {\tt emcee} runs, which is far in excess of the necessary $2\times$ the number of free parameters i.e., 12 for the Verify model, and 18 for the other three models.

\bibliography{sample631}{}
\bibliographystyle{aasjournal}
\end{document}